\documentclass[epj]{svjour}
\usepackage{graphics}
\begin{document}
\title{The Reliability on the Direction of the Incident Neutrino for the Fully Contained Events and Partially Contained Events due to QEL in the Super-Kamiokande}


\author{E. Konishi\inst{1}, Y. Minorikawa\inst{2}, V.I. Galkin\inst{3},
M. Ishiwata\inst{4}, I. Nakamura\inst{4}, N. Takahashi\inst{1}, M. Kato\inst{5}  \and A. Misaki\inst{6}    
}                     
%
%
\institute{
Graduate School of Science and Technology, Hirosaki University, Hirosaki, 036-8561, Japan    
\and Department of Science, School of Science and Engineering, Kinki University, Higashi-Osaka, 577-8502, Japan
\and Department of Physics, Moscow State University, Moscow, 119992, Russia
\and Department of Physics, Saitama University, Saitama, 338-8570, Japan
\and Kyowa Interface Science Co.,Ltd., Saitama, 351-0033, Japan
\and Advanced Research Institute for Science and Engineering, Waseda University, Tokyo, 169-0092, Japan 
\\\email{misaki@kurenai.waseda.jp}
}
\date{Received: / Revised version: March 30, 2007 }
%
\abstract{ 
In the SK analysis of the neutrino events for {\it Fully Contained Events} and {\it Partially Contained Events}
 on their zenith angle distribution, it is assumed that the zenith angle of 
the incident neutrino is the same as that of the detected charged lepton. 
In the present paper, we examine the validity of \textit{the SK assumption on the direction} of the incident neutrinos.
Concretely speaking, we analyze muon-like events due to QEL. For the purpose,
we develop {\it Time Sequential Monte Carlo Simulation} to extract the conclusion on the validity of the SK
assumption.  In our {\it Time Sequential Simulation}, we simulate every physical process concerned as exactly as possible 
without any approximation. 
From the comparison between the zenith angle distributon of the emitted muons under \textit{the SK assumption on the direction} and the 
corresponding one obtained under our {\it Time Sequential Simulation}, it is concluded that the measurement of the direction of the incident neutrino for the neutrino events occurring inside the detector in the SK analysis turns out to be unreliable, which holds irrespective 
of the existence and/or non-existence of the neutrino oscillation. 
\PACS{
      {Superkamiokande,}{QEL,} {Fully Contained Event,} 
      {Time Sequential Monte Carlo Simulation}
     } 
} 
\authorrunning{E.Konishi et. al.,}
\titlerunning{The Reliability on the Direction of Neutrino in SK}
\maketitle

\section{Introduction}
Superkamiokande have been analyzing {\it Fully Contained Events } and {\it Partially Contained Events} which are generated 
inside the detector, and {\it Upward Through Going Events} and {\it Stopping Events } which are generated outside the detector, 
for the studies on the neutrino oscillation in atmospheric neutrinos.  
   The report of oscillations between muon and tau neutrinos for
atmospheric neutrinos detected with SuperKamiokande (SK, hereafter) is
claimed to be robustly established for the following reasons:

\begin{enumerate}
\renewcommand{\labelenumi}{(\arabic{enumi})}
\item The discrimination between electrons and muons in the SK energy
range, say, several hundred MeV to several GeV, 
has been proved to be almost perfect, as demonstrated by 
calibration using accelerator beams \cite{r1}
\footnote{The SK discrimination procedure between muon and electron is constructed on the average value theory. In our opinion, discrimination procedure should be examined, taking into account  the stochastic characters of the physical processes in the neutrino events concerned. If we take this effect into account, then, for example, we give uncertainties of $3^{\circ}$ to $14^{\circ}$ in the incident direction
 of the charged lepton and uncertainties of $2m$ to $7m$ in the vertex point of the events. See, our papers \cite{r2}. However, SK give $1.8^{\circ}$ to $3.0^{\circ}$ and $0.3m$ for the same physical quantities. See, accompanied two papers.}. \\    

\item The analysis for the electron-like events and the muon-like events which give the single-ring structure in {\it Fully Contained Events} and {\it Partially Contained Events} with their zenith angle distribution, based on the well established discrimination procedure mentioned in (1),   reveals a significant deficit of muon-like events but the expected level of electron-like events. 
 It is, thus, concluded that muon neutrinos oscillate into tau neutrinos which cannot be detected due to the small geometry of SK. As the most new one, the SK collaboration published their comprehensive  paper\cite{r2}. The analysis of SK data presently yields ${\rm sin}^22\theta > 0.92$ and 
$1.5\times 10^{-3}{\rm eV}^2<\Delta m^2 < 3.4\times 10^{-3}{\rm eV}^2$ 
at 90\% confidence level.\\

\item The analysis of {\it Upward Through Going Events} and 
{\it Stopping Events}, in which the neutrino interactions occur outside
the detector, leads to similar results to (2). The charged
leptons which are produced in these categories are regarded as being
exclusively muons, because electrons have negligible probabilities to
produce such events as they lose energy very rapidly in the sorrounding rock.
Thus for these events the discrimination procedure described in (1) is not required, and, therefore, the analysis here is independent of the analysis in
(2). For these events, however, the SK group obtains the same parameters for neutrino oscillations as in (2)
\footnote{It seems strange that the experimental data with different qualities  give similarly precise results, because {\it Fully Contained Events} whose information is totally inside the detector are of higher experimental qualities compared with those of both {\it Upward Through Going Events} and {\it Stopping Events}.}.\\

\item Now, SK assert that they have found the oscillatory signature in atomospheric neutrinos from L/E analysis, which should be the ultimate evidence for the exsistence of the neutrino oscillation \cite{r3}
 Our critical examination on the L/E analysis by SK will be published elsewhere.

\end{enumerate}

 As for item(3), we have clarified that SK hardly discriminate electron( neutrino) from muon( neutrino ) in the SK manner and, instead, propose more rigorous and suitable procedure with theoretical background for the discrimination between them in the preceeding two papers.\\
 Among the neutrino events both occurring inside and outside the detector, the most robust evidence for the neutrino oscillation, if exists, should have been obtained from the analysis of both electron-like events and muon-like events in {\it Fully Contained Events}. Because (i) all neccessary informations for the physical interpretation are included in  {\it Fully Contained Events} due to their character and (ii), \\
furthermore, both electron-like events and muon-like events give the single structure image free from arbitrary interpretation with the proper electron/muon discrimination procedure.
  SK treat neutrino events whose energies cover from several hundred MeV to several GeV, if the neutrino events occur inside the detector. In this energy region, Quasi Elastic Scattering(QEL) \cite{r4} is dominant compared 
with other physical processes, such as one-pion production \cite{r5}, coherent pion production \cite{r6} and deep inelastic scattering \cite{r7}. Events due to other processes, except QEL, are not free from ambiguities due to multi-ring structure of the images.

{\bf
Therefore, SK should have analyzed the muon-like events and the electron-like events with the single ring image in {\it Fully Contained Events} 
exculusively where QEL is dominant, without utilizing poorer quality events, if SK pursue to obtain the clear cut conclusion on the neutrino oscillation
\footnote{In their analysis, they really add {\it Partially Contained Events } to the experimental data as those with the same quality under the assumption that they belong to muon-like events in {\it Fully Contained Events} to raise the statistics higher. However, such the assumption lacks in theoretical background. Furthermore, SK utilize to add multi-ring structure events which are caused by one-pion roduction, coherent pion production and deep inelastic scattering. However, the discrimination among the multi-ring structures is not so easy, which may lead the worse estimation of energies as well as directions of the events concerned.}.
}

Therefore, it is essential for us to examine single ring structure events among {\it Fully Contained Events} due to QEL which have the least ambiguities among the neutrino events concerned to obtain clear cut conclusion as for the neutrino oscillation. Here, the main concern of the present paper is devoted to the detailed analysis of the muon-like events from QEL, focusing on the direction of the incident neutrino. Situation around the corresponding electron-like event is the same as in the muon-like event. The examination on the separation of {\it Fully Contained Events} from {\it Partially Contained Events} will be discussed in subsequent papers. 

Here, it should be emphasized that the direction of the incident neutrino is assumed to be the same as that of the emitted charged lepton, i.e., the (anti-)muon or (anti-)electron, in the SK analysis of both {\it Fully Contained Events} and {\it Partially Contained Events} \cite{kajita,ishitsuka}. 
The SK {\it Detector Simulation} is to be constructed without any contradiction with \textit{the SK assumption on the direction}.

From the point of orthodoxical Monte Carlo Simulation, it seems to be unnatural for SK to impose such the assumption that the direction of the incident neutrino is the same as that of the emitted lepton ( hereafter, we call this assumption simply  "\textit{the SK assumption on the direction}") upon their {\it Detector Simulation}. Obviously, one need not any assumption on the relation between the direction of the incident neutrino and that of the emitted lepton in any sense, if we develop the Monte Carlo Simulation in a rigorous manner, which will
 be shown later in the present paper.

In order to avoid any misunderstanding toward \textit{the SK assumption on the direction} we reproduce this assumption from the original SK paper:
\\
"{\it However, the direction of the neutrino must be estimated from the reconstructed direction of the products of the neutrino interaction. In water Cherenkov detectors, the direction of an observed lepton is assumed to be 
the direction of the neutrino. Fig.\ref{fig:11} and Fig.\ref{fig:12} show the estimated correlation angle between neutrinos and leptons as a function of lepton momentum. At energies below   400 MeV/c, the lepton direction has little correlation with the neutrino direction. The correlation angle becomes smaller with increasing lepton momentum. Therefore, the zenith angle dependence of the flux as a consequence of neutrino oscillation is largely washed out below 400 MeV/c lepton momentum. With increasing momentum, the effect can be seen more clearly. } " \cite{kajita}
\footnote {It could be understood from this statement that SK justify  the validity of this assumption above 400 MeV/c. However, it is not correct, because SK put "to be proved " as the proposition. See, page 101 in their paper \cite{kajita}.}. \\

On the other hand, Ishitsuka states in his Ph.D thesis which is exclusively devoted 
into the L/E analysis of the atmospheric neutrino from Super Kamiokande as follows:
\\
\\
" {\it 8.4  Reconstruction of $L_\nu$
\vskip 2mm

Flight length of neutrino is determined from the neutrino incident zenith angle, 
although the energy and the flavor are also involved. First, the direction of neutrino is 
estimated for each sample by a different way. Then, the neutrino flight lenght is 
calclulated from the zenith angle of the reconstructed direction.
\\
\\
 8.4.1 Reconstruction of Neutrino Direction

\vspace{-2mm}

{\flushleft{\underline 
{FC Single-ring Sample}
}
}

\vspace{2mm}

The direction of neutrino for FC single-ring sample is 
simply assumed to be the same as the reconstructed direction of muon.
Zenith angle of neutrino is reconstructed as follows:
\[
\hspace{2cm}\cos\Theta^{rec}_{\nu}=\cos\Theta_{\mu} \hspace{2cm}(8.17) 
\]
,where $\cos\Theta^{rec}_{\nu}$ and $\cos\Theta_{\mu}$ are 
cosine of the reconstructed zenith angle of muon and neutrino,
respectively.} "  \cite{ishitsuka}
\footnote{It should be noticed that \textit{the SK assumption on the direction} may hold on the following possible two cases: [1] The scattering angle of the emitted lepton is so small that the effect of the scattering angle could be neglected really. However, in the present case, it could not be true from Fig. \ref{fig:1} and Fig. \ref{fig:2} and Table 1. [2] One may assert that the assumption could not hold on individual case, but it could hold statistically after accumulation of large amount of the data. However, such assertion should be verified. We verify such assumption could not hold. See, Fig. \ref{fig:11} and Fig. \ref{fig:12} in the present paper.}.\\

In our understanding, SK Monte Carlo Simulation is named usually as the  {\it Detector Simulation}. It is, however, noticed that the effect of the azimuthal angles of the emitted leptons in QEL could not be taken into account in their Simulation. As will be shown in later (see Section 3), this effect greatly influences over the final zenith angle distribution of the emitted leptons. Also, the back scattering due to QEL can not be neglected for the rigorous determination of the direction of the incident neutrino, but this effect could not be treated in the SK {\it Detector Simulation}, which is beyond the application limitaition
\footnote{SK have never clarified not only the details, but also the principle and its validity
on their Monte Carlo Simulation. We hope disclosure of their {\it Detector Simulation} for open and fair scientific discussion.}.

On the other hand, we could take into account these effects correctly in our Monte Carlo Simulation which is named as {\it Time Sequential Simulation}.    

In the present paper, we carry out the full Monte Carlo Time Sequential Simulation as exactly as possible, without \textit{the SK assumption on the direction} to clarify the problematic issue raised by SK. We carry out simulation which starts from the opposite side of the Earth to the SK detector. A neutrino sampled from the atmospheric neutrino energy spectrum at the opposite side of the Earth traverses through the medium with different densities in the interior of the Earth and penetrates finally into the SK detector where the neutrino interactions occur. The emitted energy of the individual lepton thus produced and its direction are simulated exactly based on the probability function of the cross sections concerned. 

We finally show the zenith angle distribution of the emitted leptons as well as that of the incident neutrinos are quite different from corresponding ones of the SK. This indicates that the SK assumption on the direction coud not be a reliable estimator as for the determination of the direction of the incident neutrino (See, section 5).

\section{Cross Sections of Quasi Elastic Scattering in the Neutrino 
Reaction and the Scattering Angle of Charged Leptons.}
    We examine the following reactions due to the charged current interaction
(c.c.) from QEL. \\
   \begin{eqnarray}
         \nu_e + n \longrightarrow p + e^-  \nonumber\\
         \nu_{\mu} + n \longrightarrow p + \mu^- \nonumber\\
         \bar{\nu}_e + p \longrightarrow n + e^+ \\
         \bar{\nu}_{\mu}+ p \longrightarrow n + \mu^+ \nonumber
         \label{eqn:1}
   \end{eqnarray}

The differential cross section for QEL is given as follows \cite{r6}.\\
    \begin{eqnarray}
         \frac{ {\rm d}\sigma_{\ell(\bar{\ell})}(E_{\nu(\bar{\nu})}) }{{\rm d}Q^2} = 
         \frac{G_F^2{\rm cos}^2 \theta_C}{8\pi E_{\nu(\bar{\nu})}^2}
         \Biggl\{ A(Q^2) \pm B(Q^2) \biggl[ \frac{s-u}{M^2} \biggr]
         + \nonumber \\ 
C(Q^2) \biggl[ \frac{s-u}{M^2} \biggr]^2 \Biggr\},
         \label{eqn:2}
    \end{eqnarray}

\noindent where
    \begin{eqnarray*}
      A(Q^2) &=& \frac{Q^2}{4}\Biggl[ f^2_1\biggl( \frac{Q^2}{M^2}-4 \biggr)+ f_1f_2 \frac{4Q^2}{M^2} \\
 &&+  f_2^2\biggl( \frac{Q^2}{M^2} -\frac{Q^4}{4M^4} \biggr) + g_1^2\biggl( 4+\frac{Q^2}{M^2} \biggl) \Biggr], \\
      B(Q^2) &=& (f_1+f_2)g_1Q^2, \\
      C(Q^2) &=& \frac{M^2}{4}\biggl( f^2_1+f^2_2\frac{Q^2}{4M^2}+g_1^2 \biggr).
    \end{eqnarray*}

\noindent The signs $+$ and $-$ refer to $\nu_{\mu(e)}$ and $\bar{\nu}_{\mu(e)}$ for charged current (c.c.) interactions, respectively.  The $Q^2$ denotes the four momentum transfer between the incident neutrino and the charged lepton. Details of other symbols are given in \cite{r4}.

The relation among $Q^2$, $E_{\nu(\bar{\nu})}$, the energy of the incident neutrino, $E_{\ell}$, the energy of the emitted charged lepton (muon or electron or their anti-particles) and $\theta_{\rm s}$, the scattering angle of the emitted lepton, is given as
      \begin{equation}
         Q^2 = 2E_{\nu(\bar{\nu})}E_{\ell}(1-{\rm cos}\theta_{\rm s}).
                  \label{eqn:3}
      \end{equation}

\noindent Also, the energy of the emitted lepton is given by
      \begin{equation}
         E_{\ell} = E_{\nu(\bar{\nu})} - \frac{Q^2}{2M}.
\label{eqn:4}
      \end{equation}

Now, let us examine the magnitude of the scattering angle of the emitted lepton in a quantitative way, as this plays a decisive role in determining the accuracy of the direction of the incident neutrino,
which is directly related to the reliability of the zenith angle distribution of both {\it Fully Contained Events} and {\it Partially Contained Events} in SK.

By using Eqs. (\ref{eqn:2}) to (\ref{eqn:4}), we obtain the distribution function for the scattering angle of the emitted leptons and the related quantities by a Monte Carlo method. The procedure for determining the scattering angle for a given energy of the incident neutrino is described in the Appendix A.  Fig. \ref{fig:1} shows this relation for muon, from which we can easily understand that the scattering angle $\theta_{\rm s}$ of the emitted lepton ( muon here ) cannot be neglected.  For a quantitative examination of the scattering angle, we construct the distribution function for ${\theta_{\rm s}}$ of the emitted lepton from Eqs. (\ref{eqn:2}) to (\ref{eqn:4}) by using a Monte Carlo method.

Fig. \ref{fig:2} gives the distribution function for $\theta_{\rm s}$ of the muon produced in the muon neutrino interaction. It can be seen that the muons produced from lower energy neutrinos are scattered over wider angles and that a considerable part of them are scattered even in backward directions. 
Similar results are obtained for anti-muon neutrinos, electron neutrinos and anti-electron neutrinos.
%

\begin{figure}
\begin{center}
\rotatebox{90}{%
\resizebox{0.25\textwidth}{!}{\includegraphics{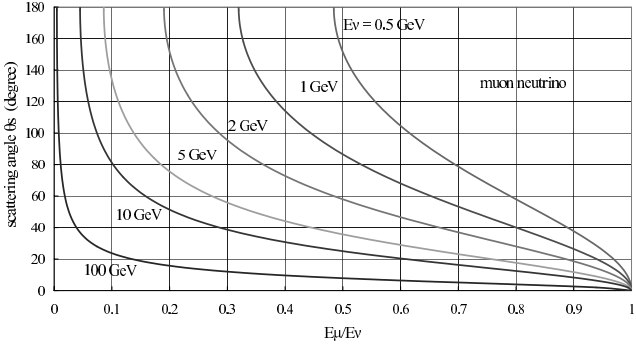}}}
\end{center}
\caption{\label{fig:1} Relation between the energy of the muon and its 
scattering angle for different incident muon neutrino energies,
 0.5~GeV, 1~GeV, 2~GeV, 5~GeV, 10~GeV and 100~GeV.}
\label{fig:1}
\end{figure} 
\begin{figure}
\rotatebox{-90}{%
\resizebox{0.3\textwidth}{!}{\includegraphics{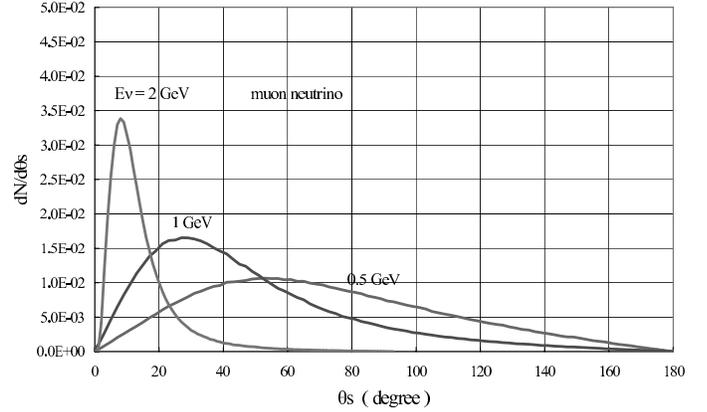}}}
\caption{\label{fig:2} Distribution functions for the scattering angle of the muon for muon-neutrino with incident energies, 0.5 GeV, 1.0 GeV and 2 GeV. Each curve is obtained by the Monte Carlo 
method (one million sampling per each curve). }
\label{fig:2}
\end{figure} 

Also, in a similar manner, we obtain not only the distribution function for the scattering angle of the charged leptons, but also their average values $<\theta_{\rm s}>$ and their standard deviations $\sigma_{\rm s}$. Table~1 shows them  for muon neutrinos, anti-muon neutrinos, electron neutrinos and anti-electron neutrinos.  In the SK analysis, it is assumed that the scattering angle of the charged
particle is zero \cite{kajita,ishitsuka}.



\begin{table*}
\caption{\label{tab:table1} The average values $<\theta_{\rm s}>$ for scattering angle of the emitted charged leptons and their standard deviations $\sigma_{\rm s}$ for various primary neutrino energies $E_{\nu(\bar{\nu})}$.}
\vspace{5mm}
\begin{center}
\begin{tabular}{|c|c|c|c|c|c|}
\hline
&&&&&\\
$E_{\nu(\bar{\nu})}$ (GeV)&angle&$\nu_{\mu(\bar{\mu})}$&$\bar{\nu}_{\mu(\bar{\mu})}$&$\nu_e$&$\bar{\nu_e}$ \\
&(degree)&&&&\\
\hline
0.2&$<\theta_\mathrm{s}>$&~~ 89.86 ~~&~~ 67.29 ~~&~~ 89.74 ~~&~~ 67.47 ~~\\
\cline{2-6}
   & $\sigma_\mathrm{s}$ & 38.63 & 36.39 & 38.65 & 36.45 \\
\hline
0.5&$<\theta_\mathrm{s}>$& 72.17 & 50.71 & 72.12 & 50.78 \\
\cline{2-6}
   & $\sigma_\mathrm{s}$ & 37.08 & 32.79 & 37.08 & 32.82 \\
\hline
1  &$<\theta_\mathrm{s}>$& 48.44 & 36.00 & 48.42 & 36.01 \\
\cline{2-6}
   & $\sigma_\mathrm{s}$ & 32.07 & 27.05 & 32.06 & 27.05 \\
\hline
2  &$<\theta_\mathrm{s}>$& 25.84 & 20.20 & 25.84 & 20.20 \\
\cline{2-6}
   & $\sigma_\mathrm{s}$ & 21.40 & 17.04 & 21.40 & 17.04 \\
\hline
5  &$<\theta_\mathrm{s}>$&  8.84 &  7.87 &  8.84 &  7.87 \\
\cline{2-6}
   & $\sigma_\mathrm{s}$ &  8.01 &  7.33 &  8.01 &  7.33 \\
\hline
10 &$<\theta_\mathrm{s}>$&  4.14 &  3.82 &  4.14 &  3.82 \\
\cline{2-6}
   & $\sigma_\mathrm{s}$ &  3.71 &  3.22 &  3.71 &  3.22 \\
\hline
100&$<\theta_\mathrm{s}>$&  0.38 &  0.39 &  0.38 &  0.39 \\
\cline{2-6}
   & $\sigma_\mathrm{s}$ &  0.23 &  0.24 &  0.23 &  0.24 \\
\hline
\end{tabular}
\end{center}
\label{tab:1}
\end{table*}

%
\section{Influence of Azimuthal Angle of Quasi Elastic Scattering over the Zenith Angle of both the Fully Contained Events and Partially Contained Events}

In the present section, we examine the effect of the azimuthal angles of the emitted leptons over their own zenith angles for given zenith angles of the incident neutrinos
\footnote{Throughout this paper, we measure the zenith angles of the emitted leptons from the upward vertical direction of the incident neutrino. Consequently, notice that the sign of our direction is oposite to that of the SK ( our $\cos\theta_{\nu(\bar{\nu})}$ = - $\cos\theta_{\nu(\bar{\nu})}$ in SK)}.

For three typical cases (vertical, horizontal and diagonal), Fig. 3 gives a schematic representation of the relationship between, $\theta_{\nu(\bar{\nu})}$, the zenith angle of the incident neutrino, and ($\theta_{\rm s}$, $\phi$) a pair of scattering angle of the emitted lepton and its azimutal angle.  

From Fig. 3(a), it can been seen that the zenith angle $\theta_{\mu(\bar{\mu})}$ of the emitted lepton is not influenced by its $\phi$ in the vertical incidence of the neutrinos $(\theta_{\nu(\bar{\nu})}=0^{\rm o})$, as it must be. From Fig. 3(b), however, it is obvious that the influence of $\phi$ of the emitted leptons on their own zenith angle is the strongest in the case of horizontal incidence of the neutrino $(\theta_{\nu(\bar{\nu})}=90^{\rm o})$. Namely, one half of the emitted leptons are recognized as upward going, while the other half is classified as downward going ones. The diagonal case ( $\theta_{\nu(\bar{\nu})}=43^{\rm o}$) is intermediate between the vertical and the horizontal. In the following, we examine the cases for vertical, horizontal and diagonal incidence of the neutrino with different energies, say, $E_{\nu(\bar{\nu})}=0.5$ GeV, $E_{\nu(\bar{\nu})}=1$ GeV and $E_{\nu(\bar{\nu})}=5$ GeV. 

\begin{figure}
\begin{center}
\resizebox{0.4\textwidth}{!}{%
  \includegraphics{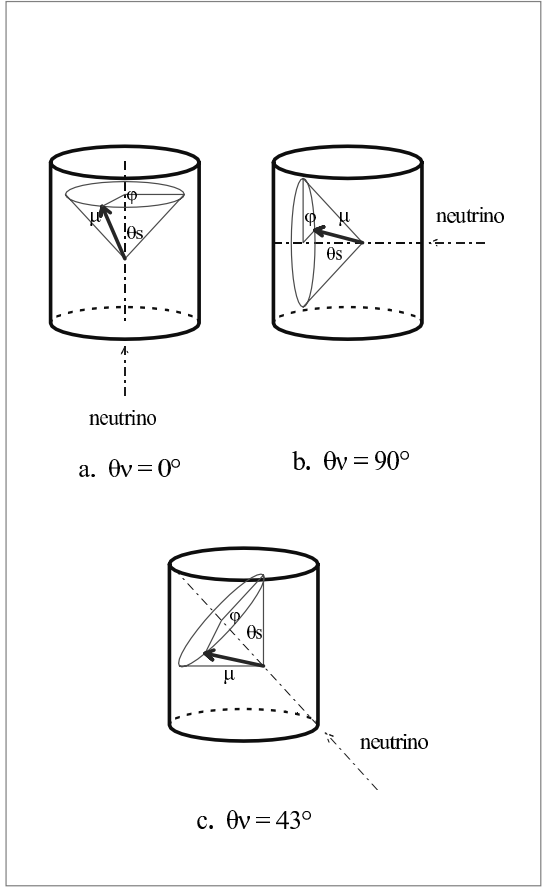}
  }
\end{center}
\caption{\label{fig:3} Schematic view of the zenith angles of the charged
 muons for diffrent zenith angles of the incident neutrinos, focusing on
  their azimuthal angles.}
\label{fig:3}
\end{figure} 
  The detailed procedure for the Monte Carlo simulation is described in the Appendix A.

\subsection{Dependence of the spreads of the zenith angle for the emitted leptons on the energies 
of the emitted leptons for different incident directions with different energies}
\begin{figure*}
\hspace{2.9cm}(a)
\hspace{5.5cm}(b)
\hspace{5.5cm}(c)
\vspace{-0.3cm}
\begin{center}
\resizebox{\textwidth}{!}{%
  \includegraphics{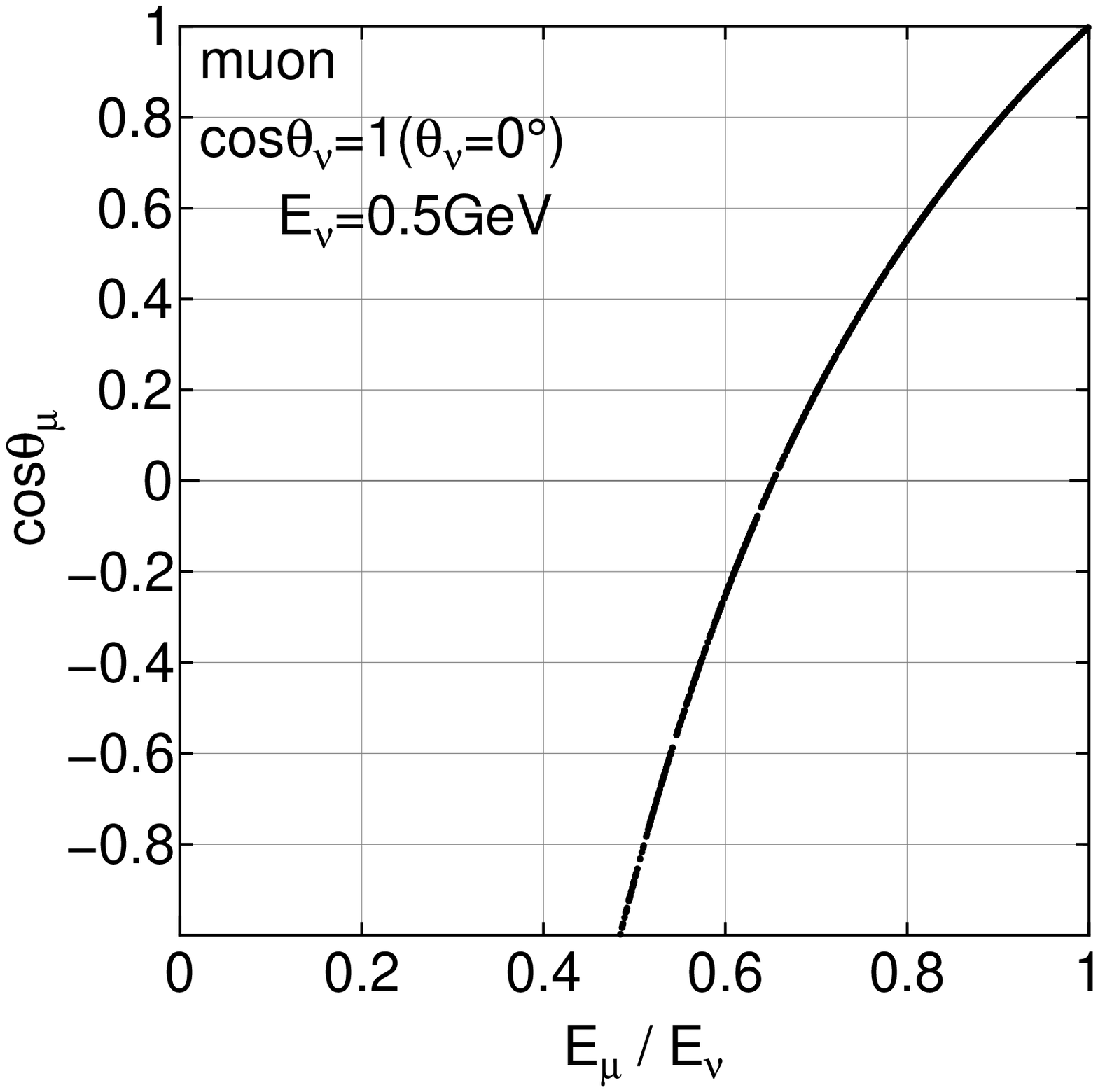}\hspace{1cm}
  \includegraphics{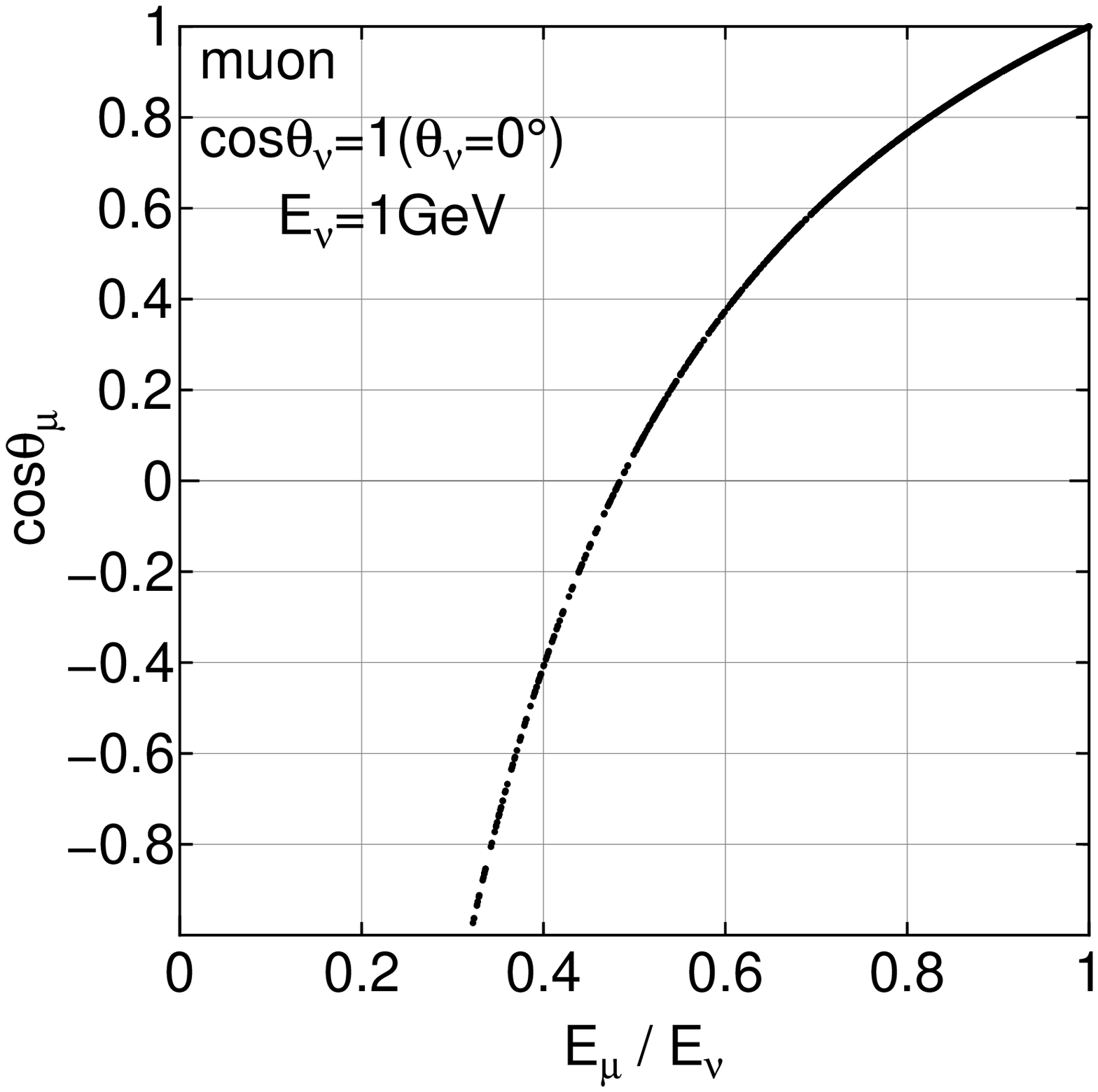}\hspace{1cm}
  \includegraphics{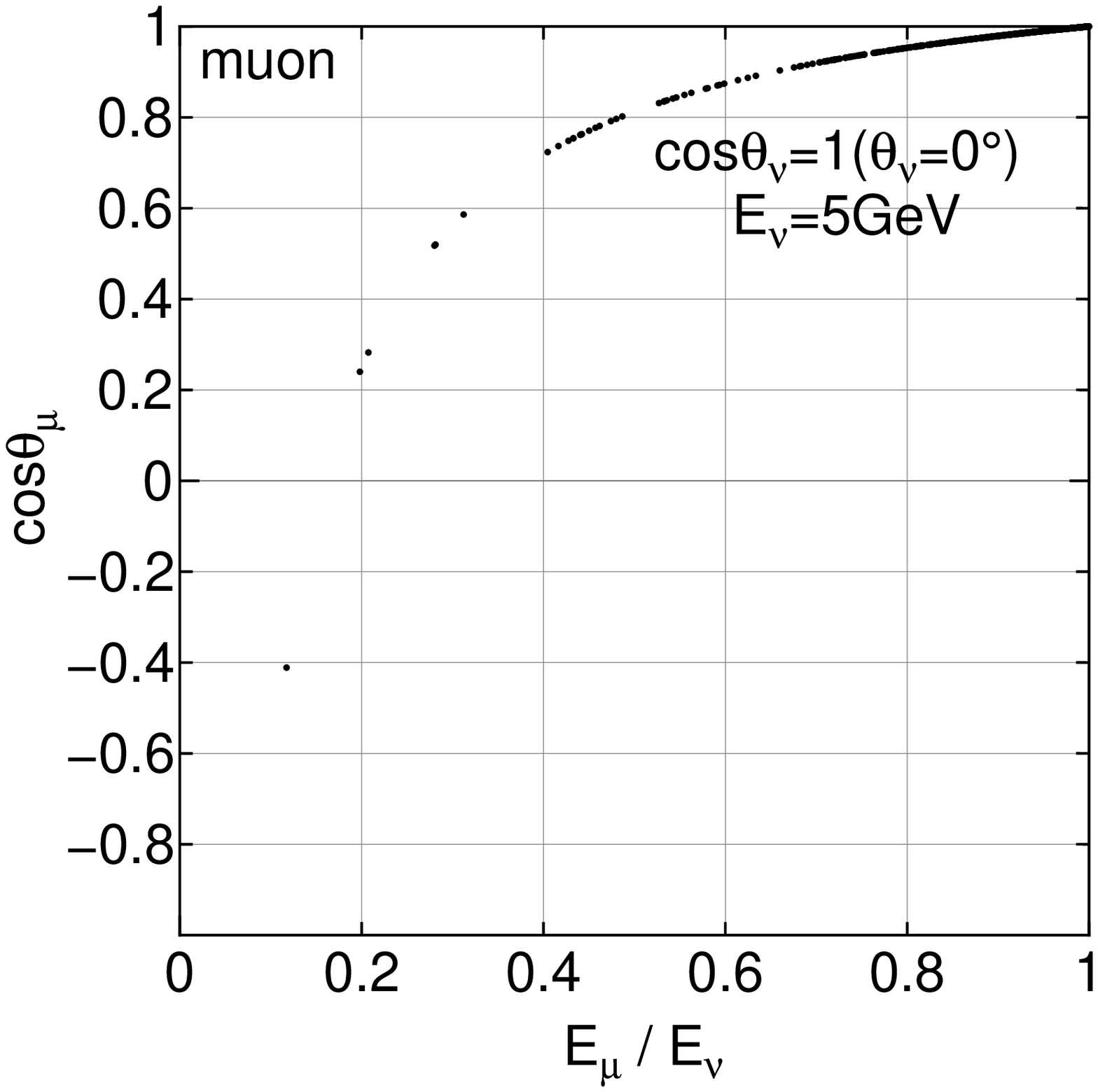}
}
\caption{
\label{fig:4}
The scatter plots between the fractional energies of the produced muons and their zenith angles 
for vertically incident muon neutrinos with 0.5~GeV, 1~GeV and 5~GeV, respectively.
 The sampling number is 1000 for each case.
}
\end{center}
\vspace{0.5cm}
\end{figure*}

\begin{figure*}
\hspace{2.9cm}(a)
\hspace{5.5cm}(b)
\hspace{5.5cm}(c)
\vspace{-0.3cm}
\begin{center}
\resizebox{\textwidth}{!}{%
  \includegraphics{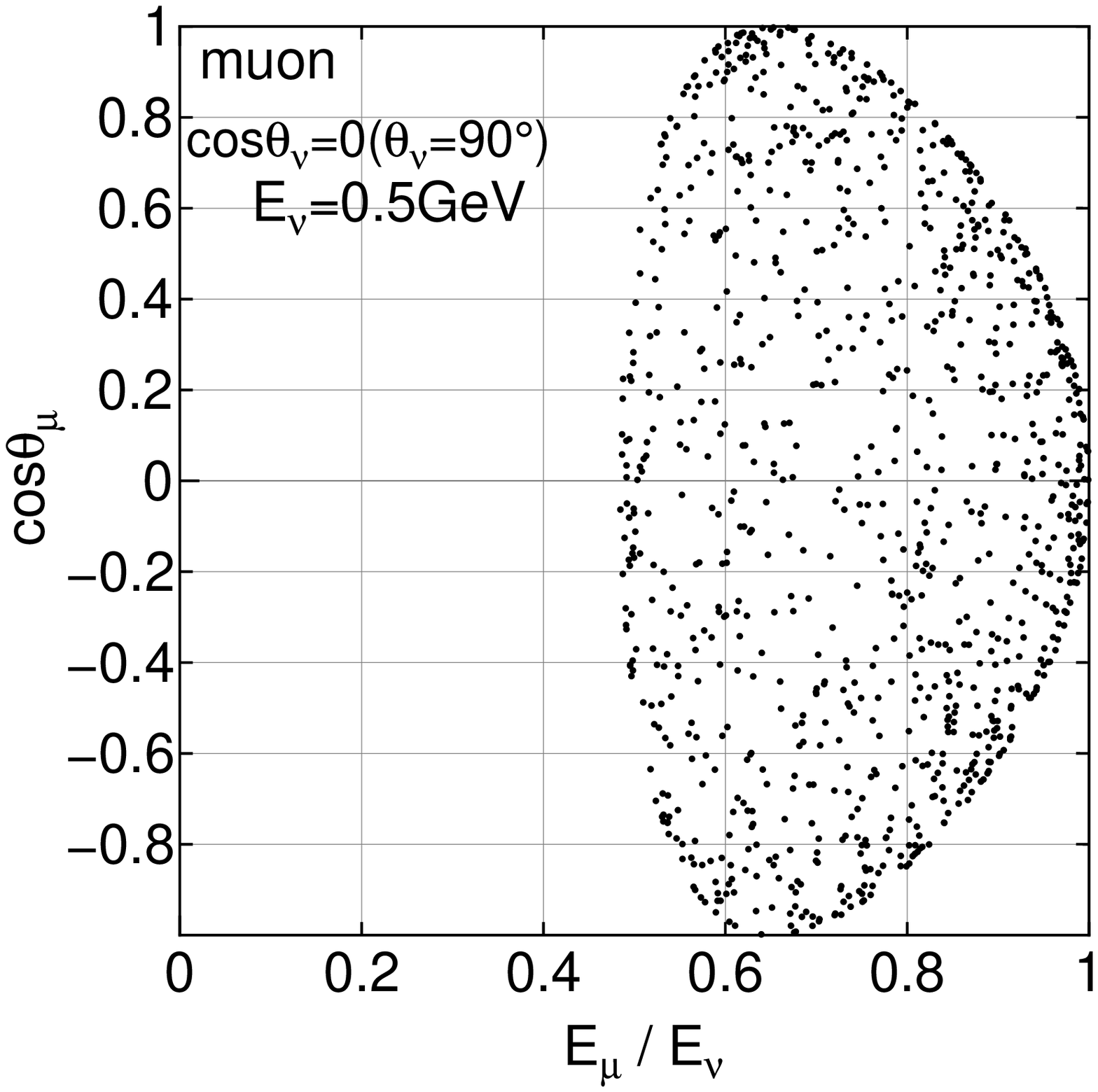}\hspace{1cm}
  \includegraphics{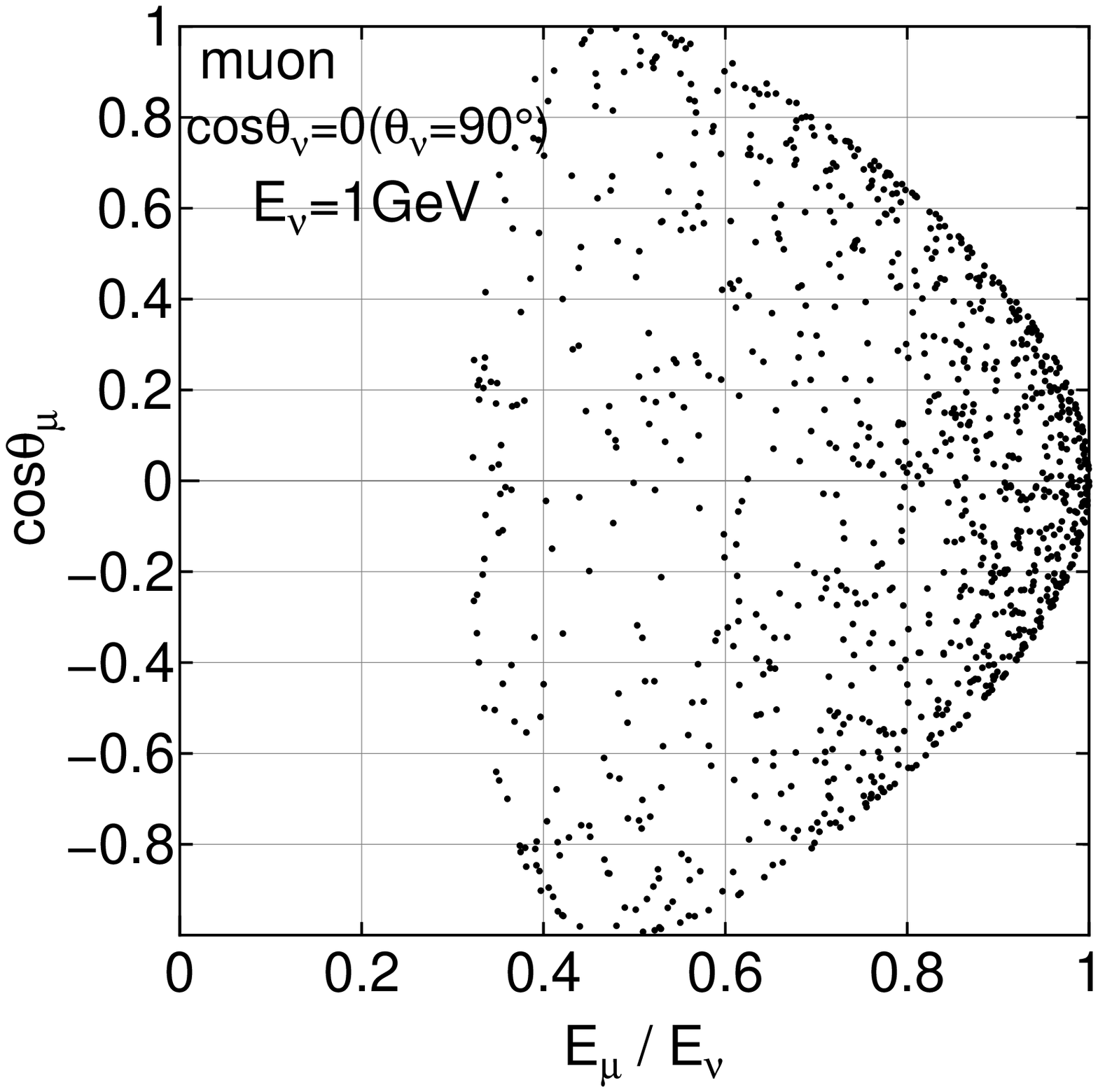}\hspace{1cm}
  \includegraphics{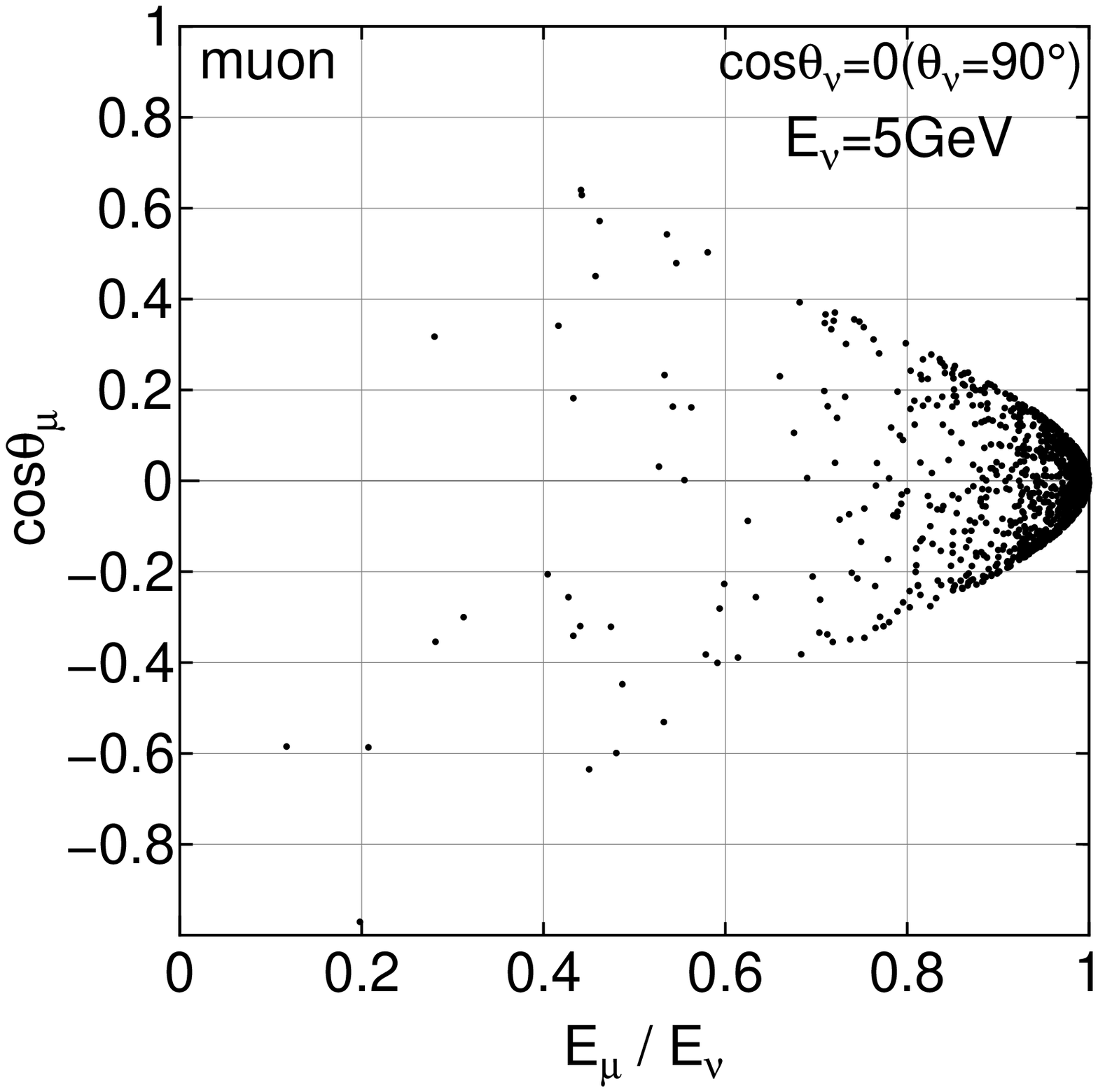}
}
\caption{
\label{fig:5} 
The scatter plots between the fractional energies of the produced muons and their zenith angles 
for horizontally incident muon neutrinos with 0.5~GeV, 1~GeV and 5~GeV, respectively.
 The sampling number is 1000 for each case.
}
\end{center}
\vspace{0.5cm}
\end{figure*}

\begin{figure*}
\hspace{2.9cm}(a)
\hspace{5.5cm}(b)
\hspace{5.5cm}(c)
\vspace{-0.3cm}
\begin{center}
\resizebox{\textwidth}{!}{%
  \includegraphics{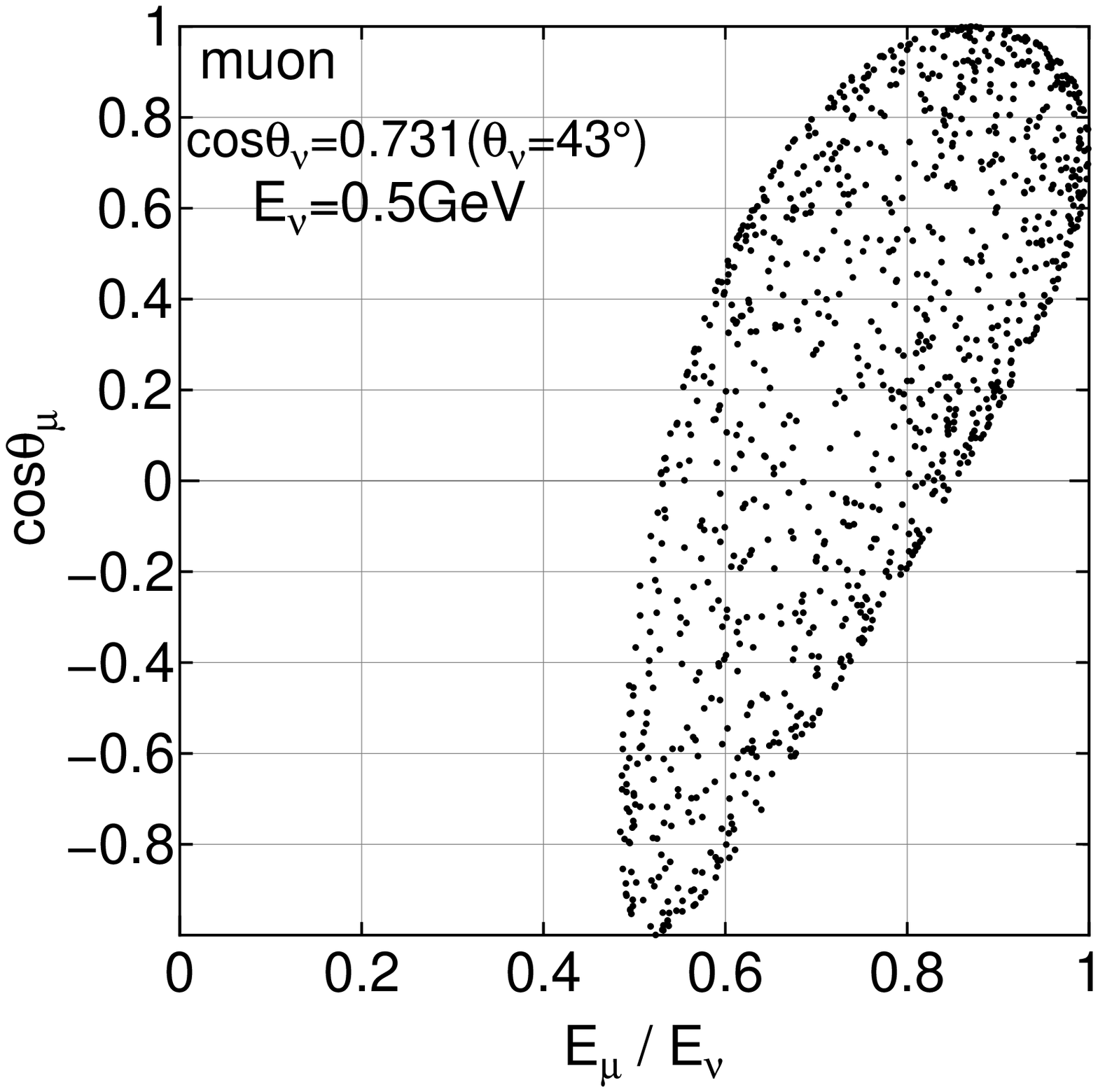}\hspace{1cm}
  \includegraphics{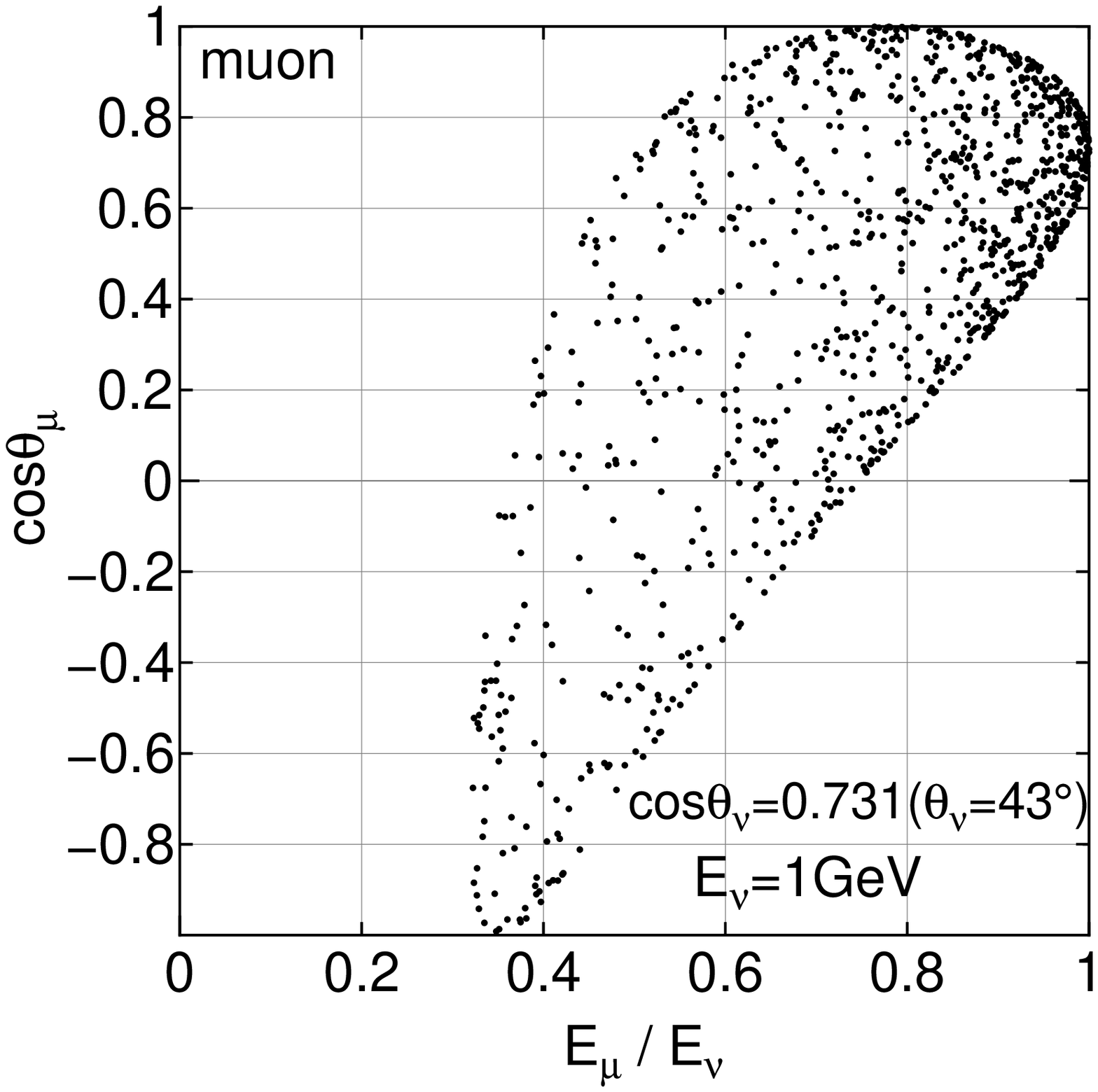}\hspace{1cm}
  \includegraphics{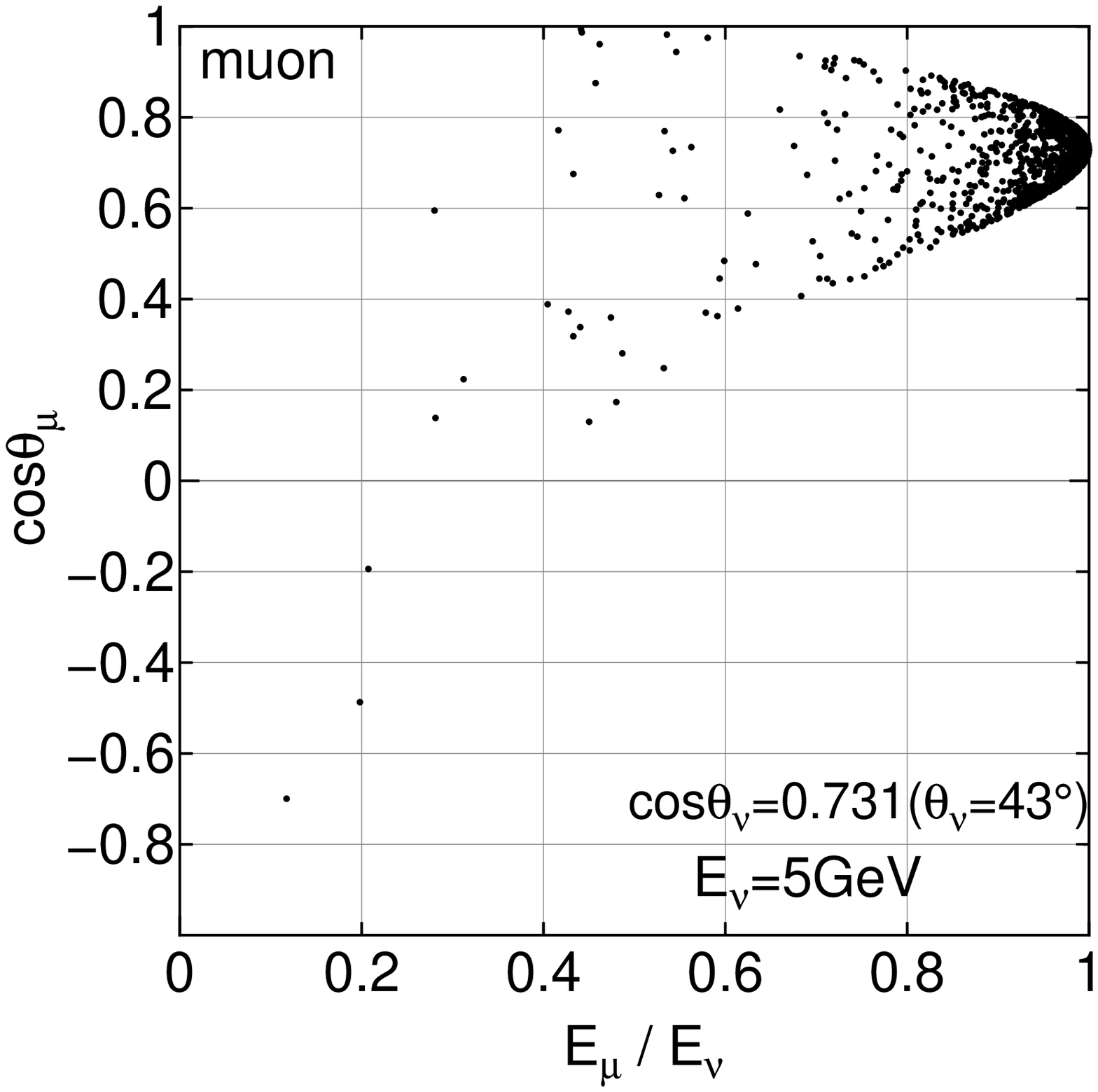}
  }
\caption{
\label{fig:6} 
The scatter plots between the fractional energies of the produced muons and their zenith angles 
for diagonally incident muon neutrinos with 0.5~GeV, 1~GeV and 5~GeV, respectively.
 The sampling number is 1000 for each case.
}
\end{center}
\end{figure*}

 We give  the scatter plots between the fractional energies of the emitted muons and their zenith angle
 for a definite zenith angles of the incident neutrino with different energies in Figs. \ref{fig:4} to \ref{fig:6}. In Fig. \ref{fig:4}, we give the scatter plots for vertically incident neutrino with different energies 0.5 GeV, 1 GeV and 5 GeV . In this case, the relations between the emitted energies of the muon and and their zenith angles are unique, which comes from the definition of the zenith angle of the emitted lepton. However, the densities (frequencies of event number) along each curve is different in position to position and depend on the energies of the incident neutrinos. Generally speaking,densities along curves become higher toward  $\cos\theta_{\mu(\bar{\mu})}= 1$. In this case, $\cos\theta_{\mu(\bar{\mu})}$ is never influenced by the azimuthal angel in the scattering by the definition
\footnote{The zenith angles of the particles concerned are measured from the vertical direction.}.
 
Fig. 5 tells us that the horizontally incident neutrinos give the most widely spread of the zenith angle distribution of the emitted lepton influenced by the azimuthal angle. The more lower incident neutrino energies, the more wider spreads of the emitted leptons. The diagonally incident neutrinos give the intermediate distribution of the emitted leptons between those of vertically incident neutrinos and horizontally incident neutrinos.

\subsection{Zenith angle distribution of the emitted lepton 
for the different incidence of the neutrinos with different energies}

\begin{figure*}
\hspace{2.9cm}(a)
\hspace{5.5cm}(b)
\hspace{5.5cm}(c)
\vspace{-0.3cm}
\begin{center}
\resizebox{\textwidth}{!}{%
  \includegraphics{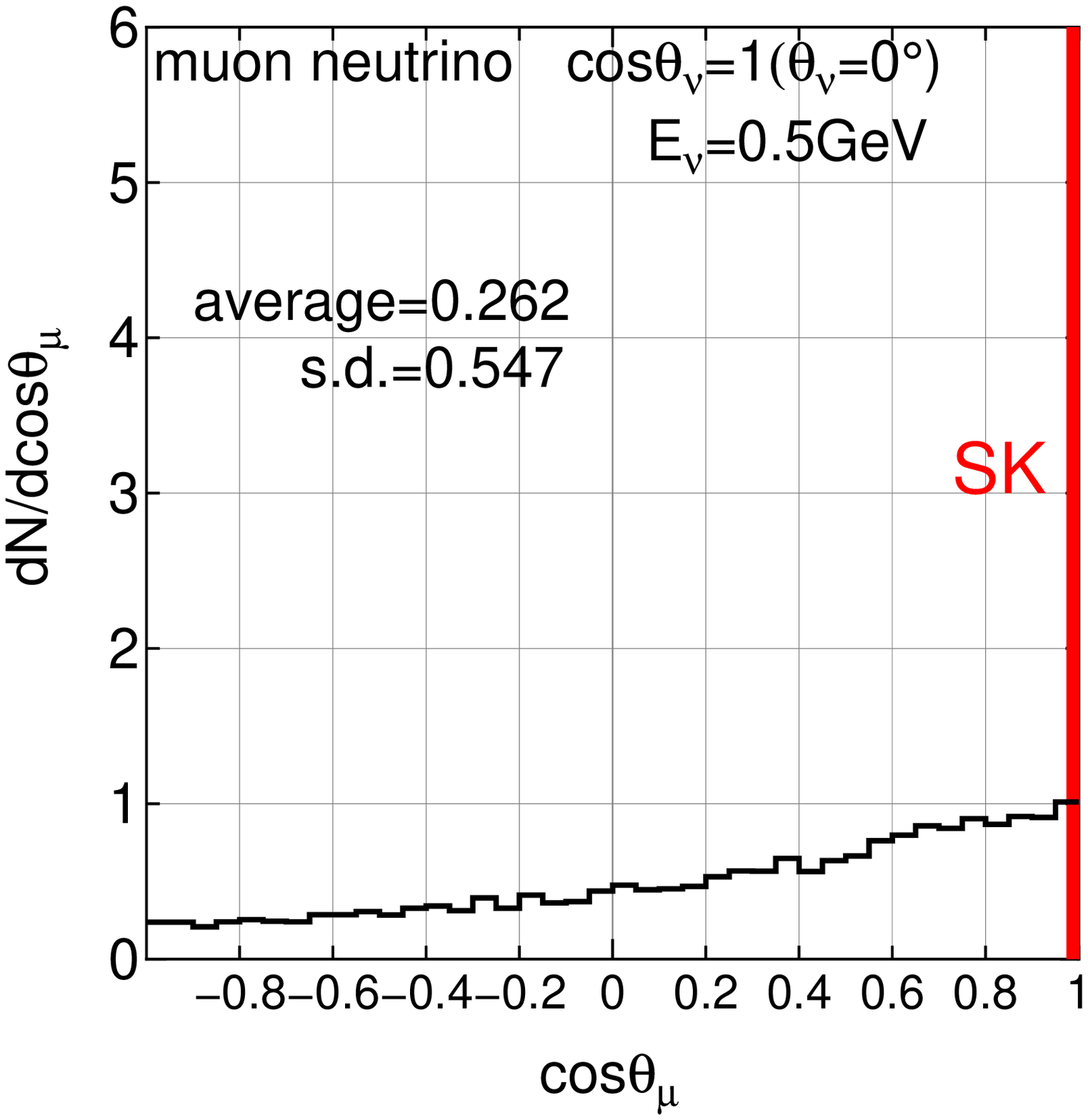}\hspace{1cm}
  \includegraphics{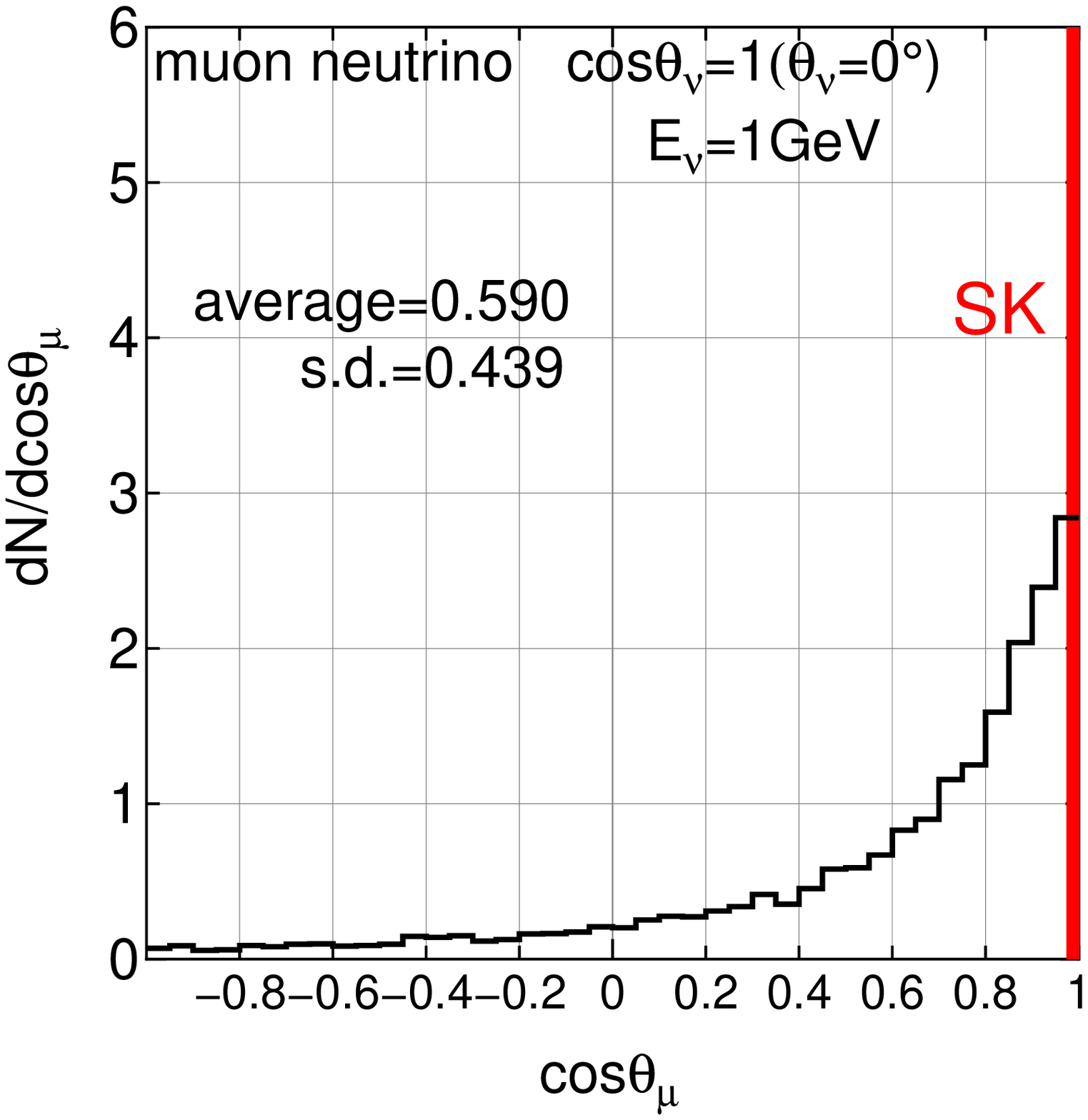}\hspace{1cm}
  \includegraphics{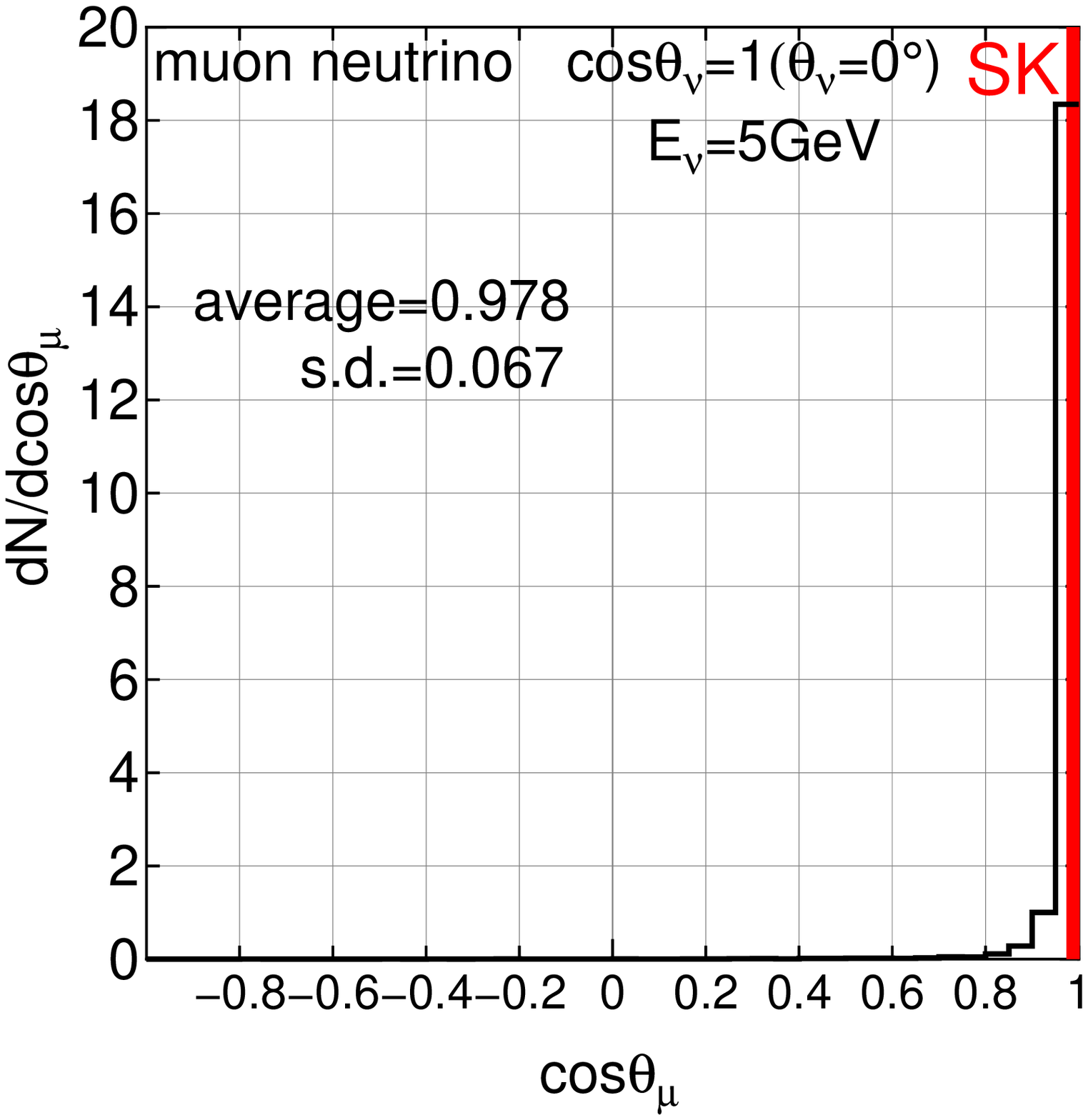}
  }
\caption{
\label{fig:7} 
Zenith angle distribution of the muon for the vertically incident muon neutrino 
with 0.5~GeV, 1~GeV and 5~GeV, respectively. The sampling number is 10000 for each case.
SK stand for the corresponding ones under the SK assumption.
}
\end{center}
\vspace{0.5cm}
\end{figure*} 

\begin{figure*}
\hspace{2.9cm}(a)
\hspace{5.5cm}(b)
\hspace{5.5cm}(c)
\vspace{-0.3cm}
\begin{center}
\resizebox{\textwidth}{!}{%
 \includegraphics{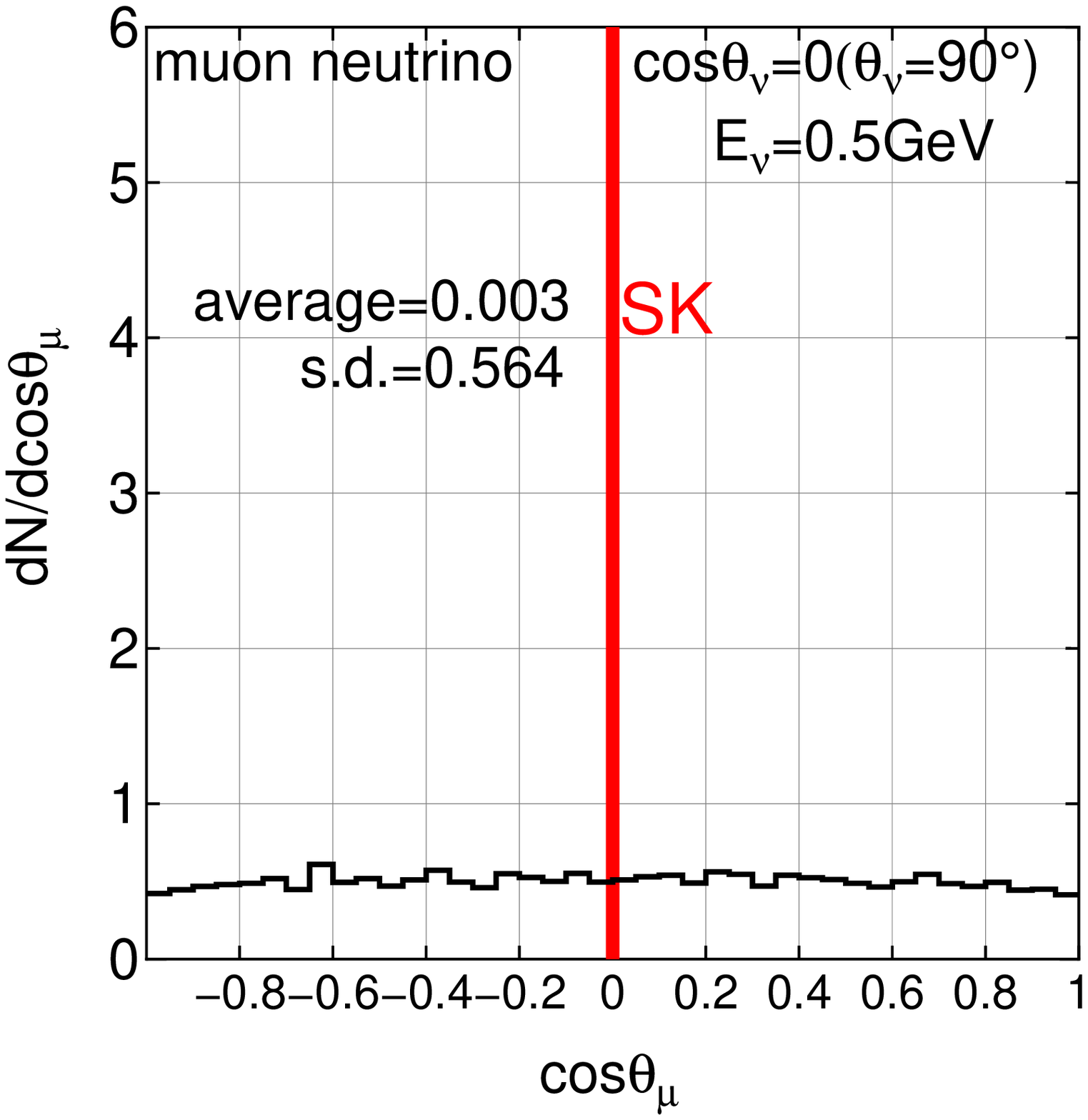}\hspace{1cm}
  \includegraphics{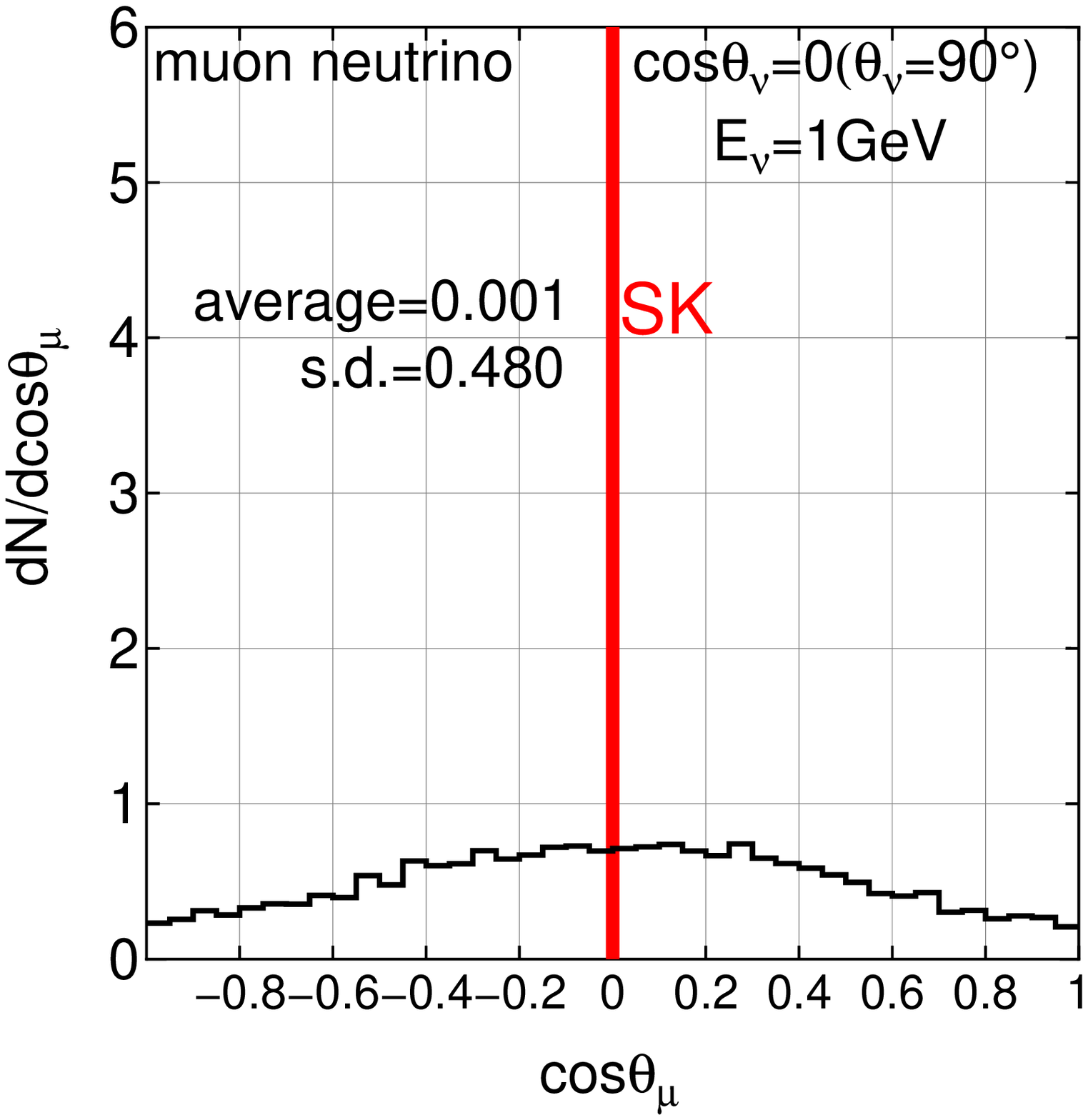}\hspace{1cm}
  \includegraphics{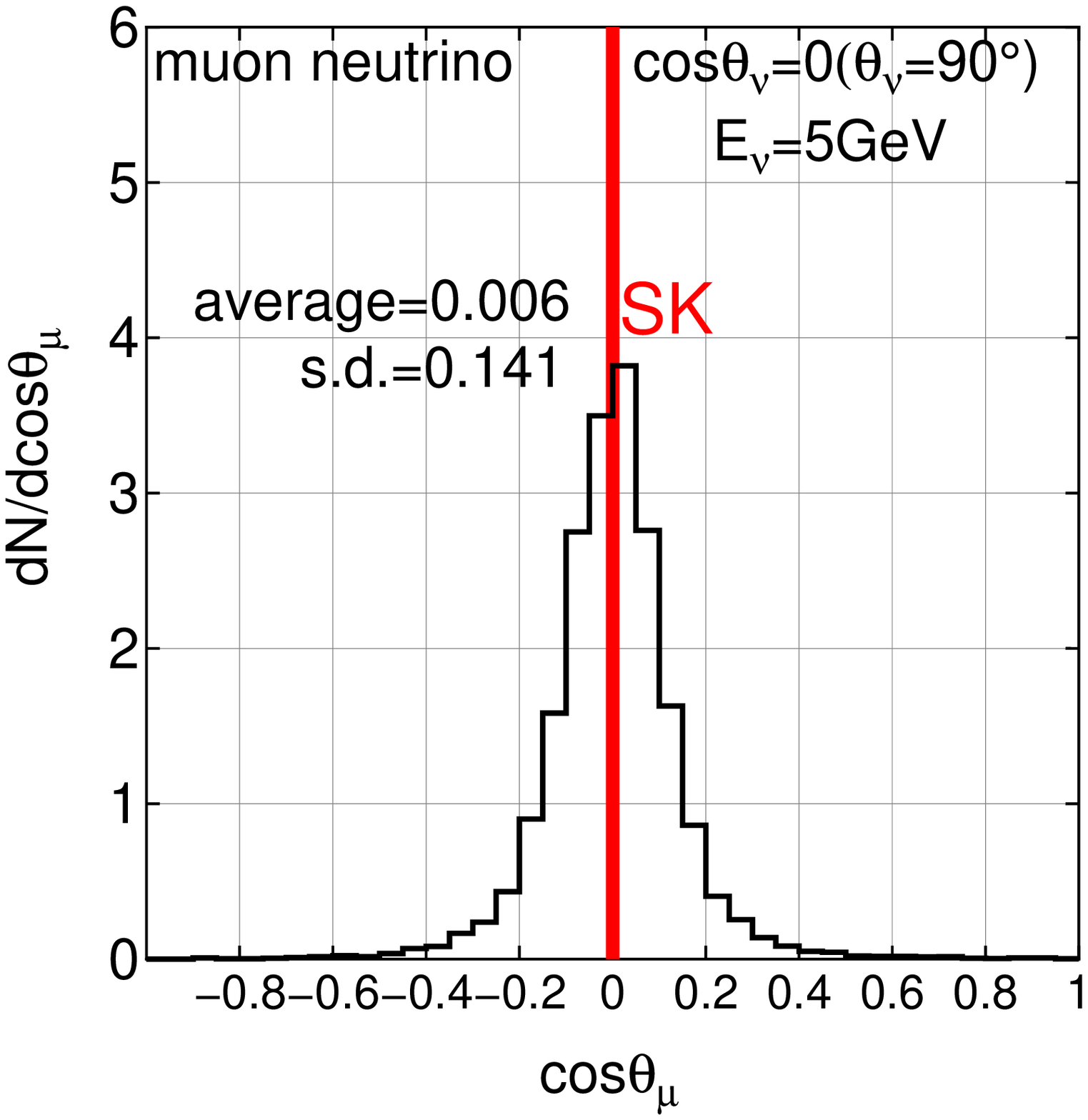}
  }
\caption{
\label{fig:8} 
Zenith angle distribution of the muon for the horizontally incident muon neutrino 
with 0.5~GeV, 1~GeV and 5~GeV, respectively. The sampling number is 10000 for each case.
SK stand for the corresponding ones under the SK assumption.
}
\end{center}
\vspace{0.5cm}
\end{figure*} 
\begin{figure*}
\hspace{2.9cm}(a)
\hspace{5.5cm}(b)
\hspace{5.5cm}(c)
\vspace{-0.3cm}
\begin{center}
\resizebox{\textwidth}{!}{%
  \includegraphics{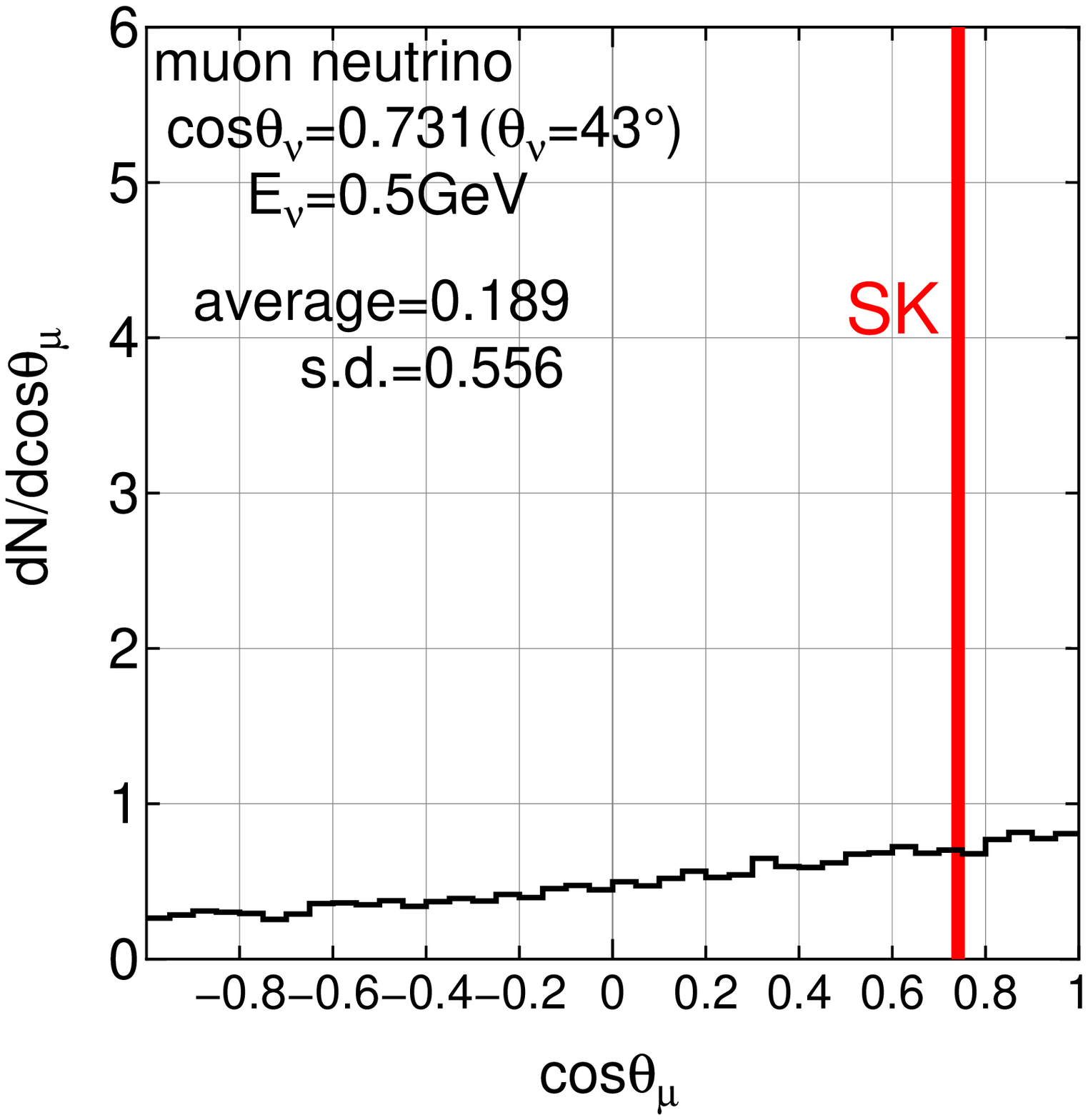}\hspace{1cm}
  \includegraphics{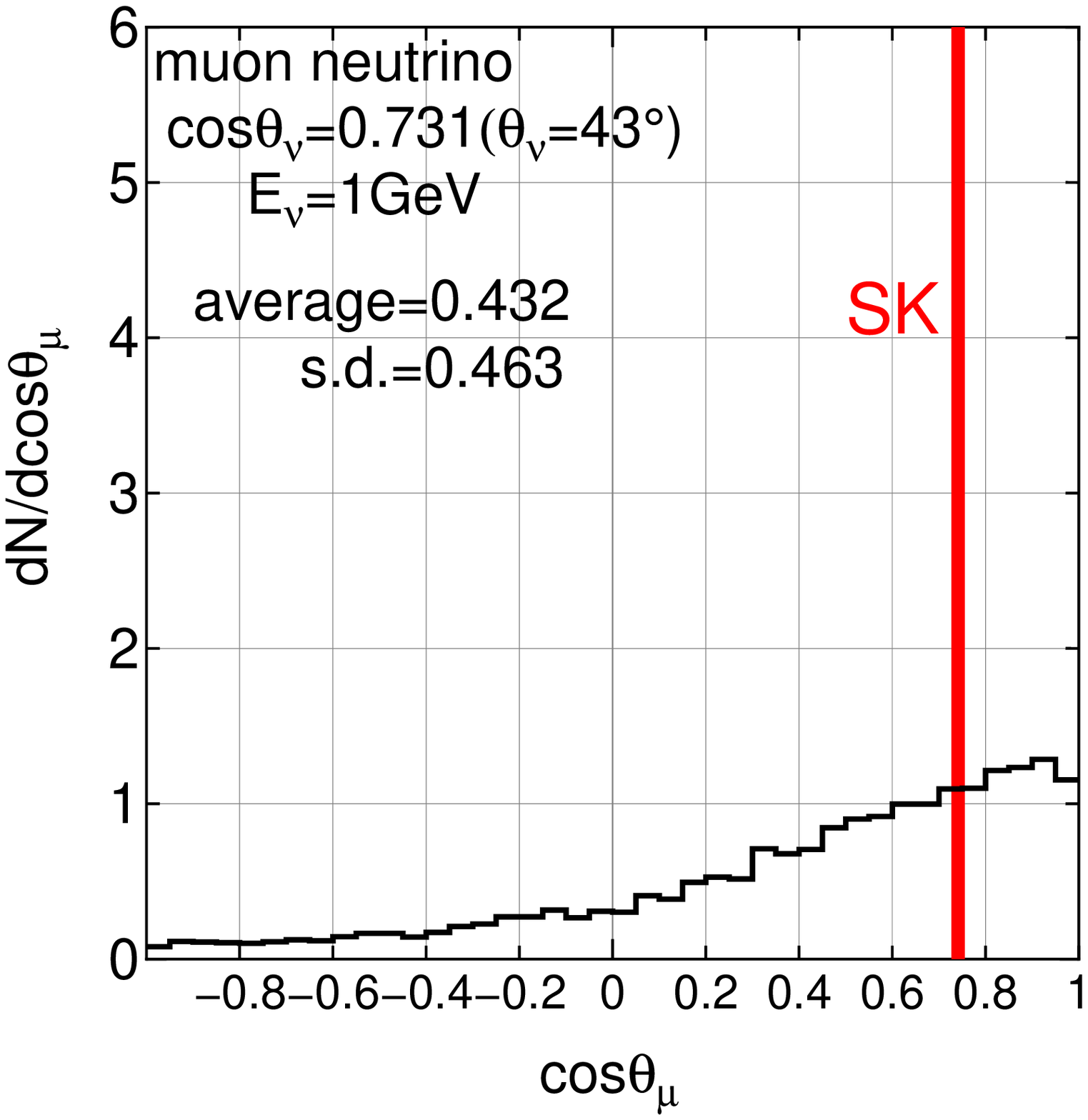}\hspace{1cm}
  \includegraphics{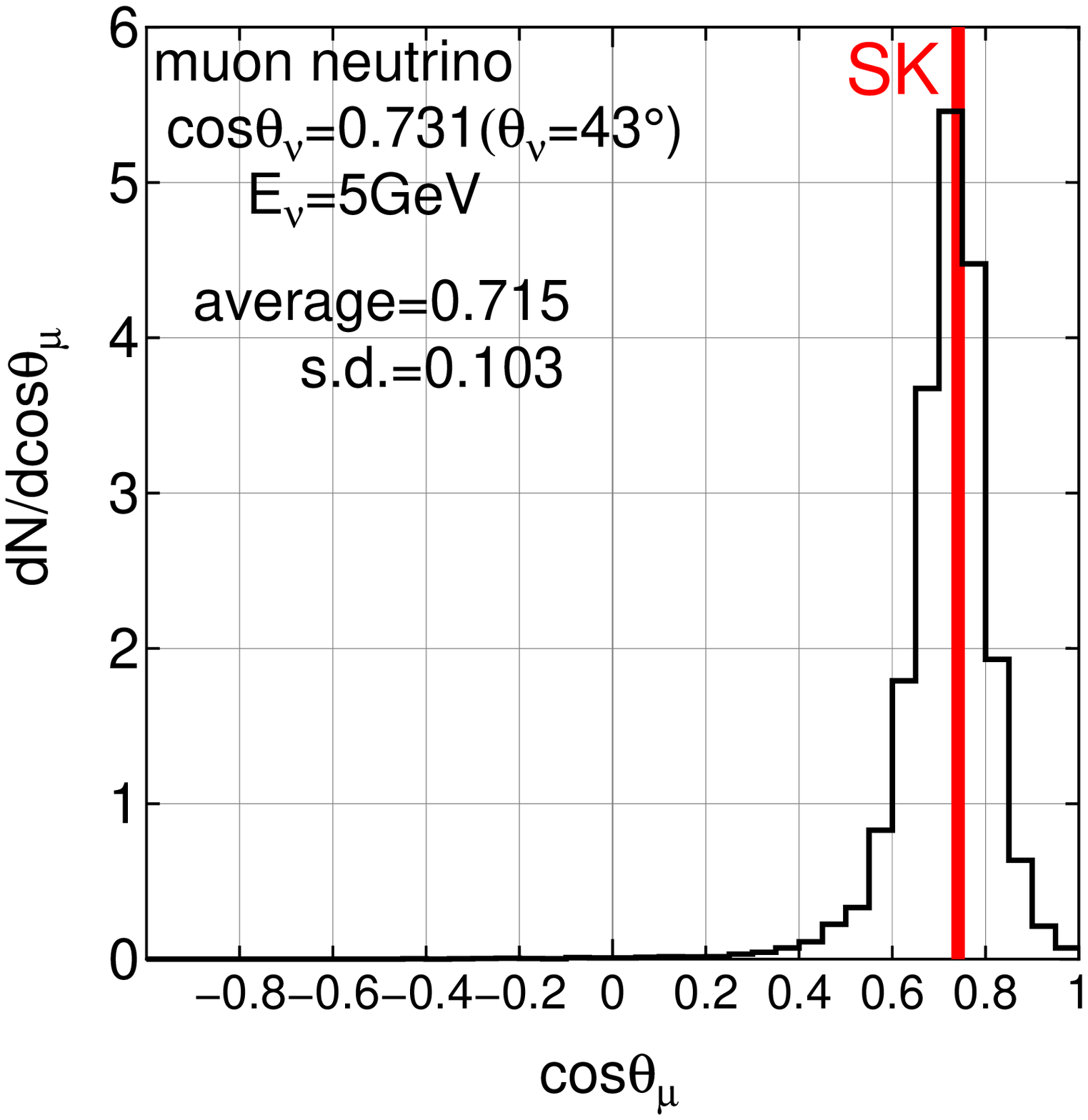}
  }
\caption{
\label{fig:9} 
Zenith angle distribution of the muon for the diagonally incident muon neutrino 
with 0.5~GeV, 1~GeV and 5~GeV, respectively. The sampling number is 10000 for each case.
SK stand for the corresponding ones under the SK assumption.
}
\end{center}
\end{figure*}

In Figs. \ref{fig:7} to \ref{fig:9}, we give the zenith angle distributions of the emitted muons for the given direction of the incident neutrinos with different energies of the neutrino. These figures are obtained through summation on the energies of the emitted muons for their definite zenith angles in Figs. \ref{fig:4} to \ref{fig:6}.

In Figs. \ref{fig:7}(a) to \ref{fig:7}(c), we give the zenith angle distribution of the emitted muon for the case of vertically incident neutrinos with different energies, say, $E_{\nu}=0.5$ GeV, $E_{\nu}=1$ GeV and $E_{\nu}=5$ GeV.

Comparing the case for 0.5 GeV with that for 5 GeV, we understand the big contrast between them as for the zenith angle distribution. The scattering angle of the emitted muon for 5 GeV neutrino is relatively small (See, Table 1) that the emitted muons keep roughly the same direction as their original neutrino. In this case, the effect of their azimuthal angle on the zenith angle is also small. However, in the case for 0.5 GeV which is  the dominant energy for \textit{Fully Contained Events} in the Superkamiokande, there is even 
a possibility for the emitted muon to be emitted in the backward direction 
due to the large angle scattering, the effect of which is enhanced by their azimuthal angle.

The most frequent occurrence in the backward scattering of the emitted muon appear in the horizontally incident neutrino as shown in Figs. 8(a) to 8(c). In this case, the zenith angle distribution of the emitted muon should be symmetrical to the horizontal direction. Comparing the case for 5 GeV with those both for $\sim$0.5 GeV and for $\sim$1 GeV, even 1 GeV incident neutrinos lose almost the original sense of the incidence if we measure it by the zenith angle of the emitted muon. Figs. 9(a) to 9(c) for the diagonally incident neutrino tell us that the situation for diagonal cases lies between the case for the vertically incident neutrino and that for horizontally incident one.

\section{Zenith Angle Distribution of Fully Contained Events and Partially Contained Events 
for a Given Zenith Angle of the Incident Neutrino, Taking Their Energy Spectrum into Account}

\begin{table*}
\caption{\label{tab:table2}Average values and their standerd deviations in cos${\theta}_{\mu+\bar{\mu}}$ for the zenith angle distributions of the muons with different primary energies of the insident neutrinos.}
\begin{center}
\begin{tabular}{c c c c c c c }
\hline
& \multicolumn{2}{c}{Vertical}  & \multicolumn{2}{c}{Diagonal}  & \multicolumn{2}{c}{Horizontal} \\
& \multicolumn{2}{c} { cos$\theta_{\nu+\bar{\nu}}=1\, (0^{\circ})$} &  \multicolumn{2}{c} 
{cos$\theta_{\nu+\bar{\nu}}=0.731\, (43^{\circ}) $} & \multicolumn{2}{c}{ cos$\theta_{\nu+\bar{\nu}}=0\, (90^{\circ}) $} \\
\hline
$E_{\nu+\bar{\nu}}$(GeV) & $\overline{\cos\theta_{\mu+\bar{\mu}}}$ &  $\sigma_{\cos\theta_{\mu+\bar{\mu}}}$  &\, \,  $\overline{\cos\theta_{\mu+\bar{\mu}}} $ & $\sigma_{\cos\theta_{\mu+\bar{\mu}}}$ & 
\, $\overline{\cos\theta_{\mu+\bar{\mu}}} $ & $\sigma_{\cos\theta_{\mu+\bar{\mu}}} $ \\
\hline
 0.5 & 0.262 & 0.547 &\, \,  0.189 & 0.556 &\,  -0.003 & 0.564 \\
1.0 & 0.590 & 0.439 &\, \,  0.432 & 0.463 &\,  0.001 & 0.480 \\
2.0 & 0.581 & 0.250 &\, \,  0.623 & 0.290 &\,  0.001 & 0.325 \\
5.0 & 0.978 & 0.067 &\, \,  0.715 & 0.103 &\,  0.006 & 0.141 \\
\hline
Spectrum$^{*}$ & 0.468 & 0.531 &\, \,  0.339 & 0.519 &\,  -0.005 & 0.500 \\
SK$^{**}$ & 1.00 &  0.00 &\, \,  0.731 & 0.000  &\,  0.000 & 0.000 \\
\hline
\end{tabular}
\end{center}
\end{table*}

In the previous sections, we discuss the relation between the zenith angle distribution of the incident neutrino with a single energy and that of the emited muons produced by the neutrino for the different incident direction. In order to apply our motivation  around the uncertainty of \textit{the SK assumption on the direction} for \textit{Fully Contained Events} and \textit{Partially Contained Events}, we must consider the effect of the energy spectrum of the incident neutrino. The Monte Carlo simulation procedure for this purpose are given in the Appendix B.

\begin{figure*}
\hspace{2.9cm}(a)
\hspace{5.5cm}(b)
\hspace{5.5cm}(c)
\vspace{-0.5cm}
\begin{center}
\resizebox{\textwidth}{!}{%
  \includegraphics{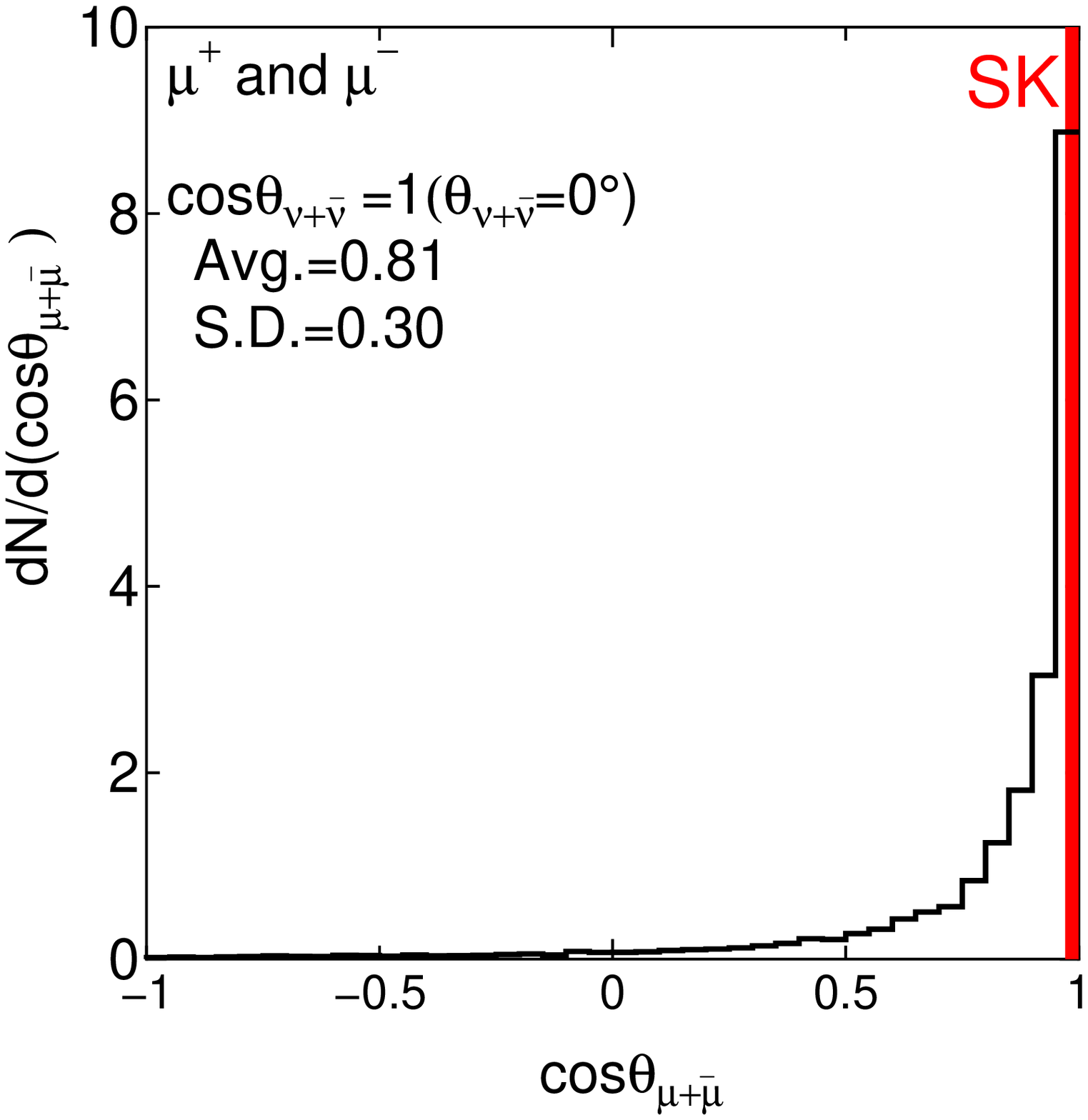}\hspace{1cm}
  \includegraphics{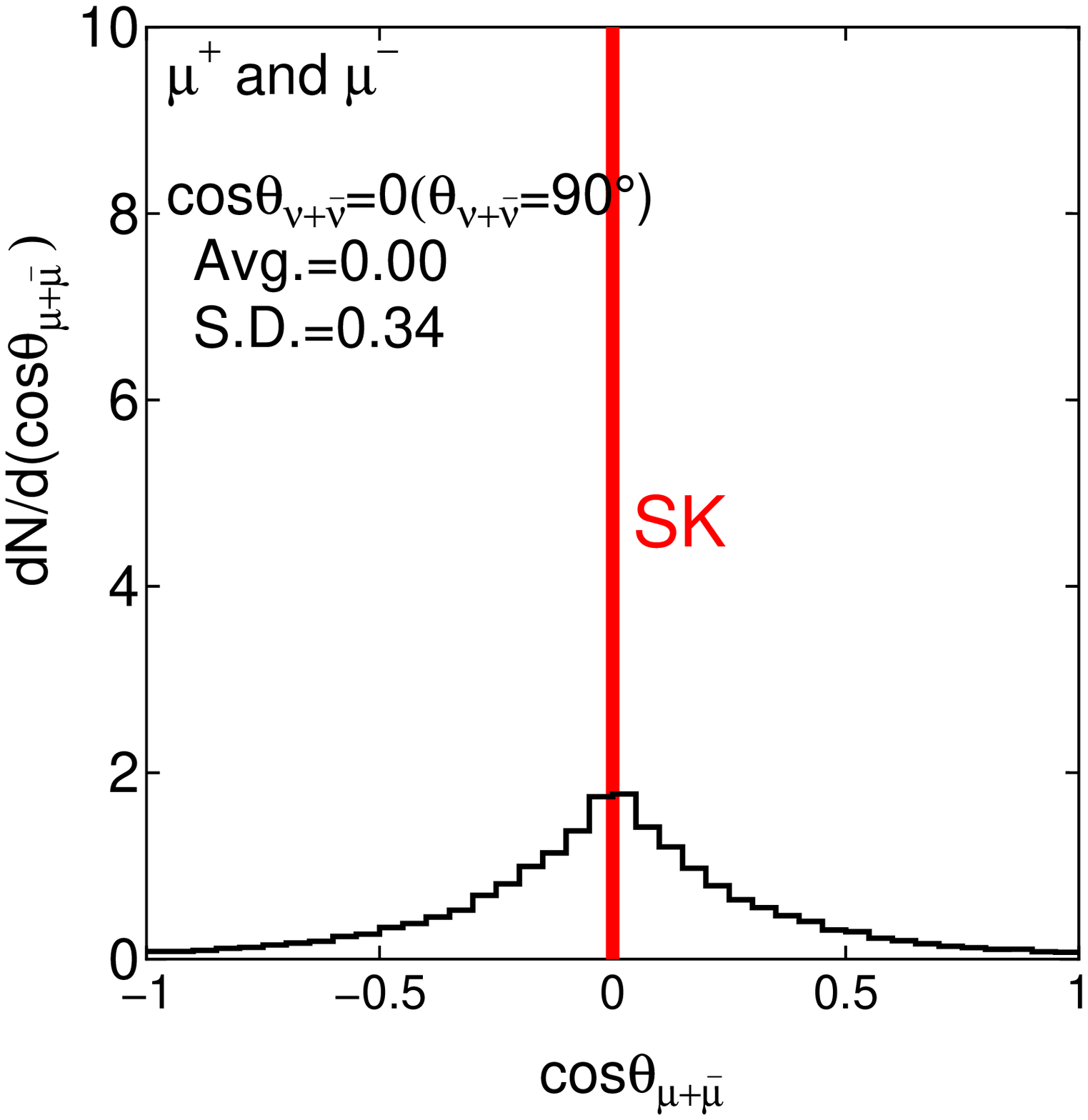}\hspace{1cm}
  \includegraphics{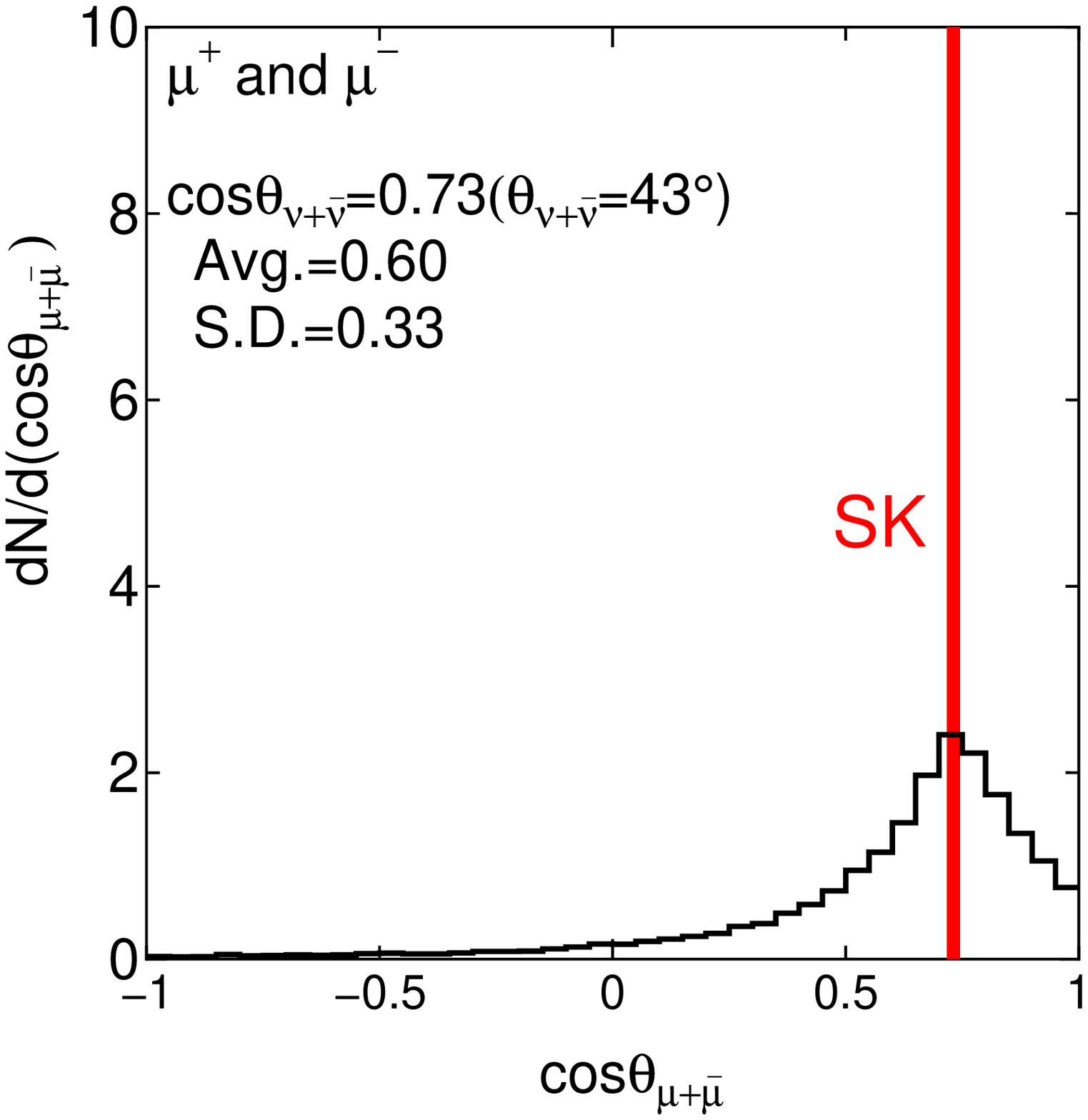}\hspace{1cm}
  }\\
 \end{center}
\caption{\label{fig:10} 
Zenith angle distribution of $\mu^-$ and $\mu^+$ for  $\nu$  and   $\bar{\nu}$ for the incident neutrinos with the vertical, horizontal and diagonal directions, respectively. The overall neutrino spectrum at Kamioka site is taken into account. The sampling number is 10000 for each case. SK stand for the corresponding ones under the SK assumption.
}
\end{figure*}

In Fig. \ref{fig:10}, we give the zenith angle distributions of the sum of $\mu^+(\bar{\mu})$ and $\mu^-$ for a given zenith angle of $\bar{\nu}_{\bar{\mu}}$ and $\nu_{\mu}$, taking into account
primary neutrino energy spectrum at Kamioka site.

In Table 2, the average values for ${\rm cos}{\theta}_{\mu+\bar{\mu}}$ and their standard deviation for different incidences of the incident neutrinos with different energies are presented
\footnote{Notice that the difference in the corresponding quantites between the case for single energy and the case for the energy spectrum. The formers are given in the $\mu^-$,while th latter is given in $\mu^-$ and $\mu^+$. However, such the difference does not change the essential recognition.}.

In the SK case, their average values are given by ${\rm cos}{\theta}_{\nu(\bar{\nu})}$ themselves by definition and, consequently, the standard deviations are zero under the assumption, because the SK assumption is of the delta function for the incidence direction. They are shown in the bottom line of Table 2.
In the second line from the bottom in this table, we give the average values and their standard deviations for ${\cos\theta_{\nu+\bar{\nu}}}$ obtained under the inclusion of the energy spectrum for primary neutrinos. Thus, we found these values correspond to those for incident neutrino with the effective single energy between 0.5 GeV and 1 GeV. If we compare  the average energies and the standard deviations for the inclusion of incident neutrino energy spectrum with those under the SK assumption, it is easily understood that SK assumption does not represent real zenith angle distribution of the emitted muon.
 

\section{Relation between the Zenth angle Distribution of the Incident Neutrinos and that of the emitted leptons}

Now, we extend the results for the definite zenith angle obtained in the previous section to the case in which we consider the zenih angle distribution of the incident neutrinos totally.

Here, we examine the real correlation between ${\cos\theta_{\nu}}$ and ${\cos{\theta}_{\mu}}$, by peforming the exact Monte Carlo simulation. 
 

The detail for the simulation procedure is given the Appendix C.

\begin{figure}
\begin{center}
\resizebox{0.45\textwidth}{!}{%
  \includegraphics{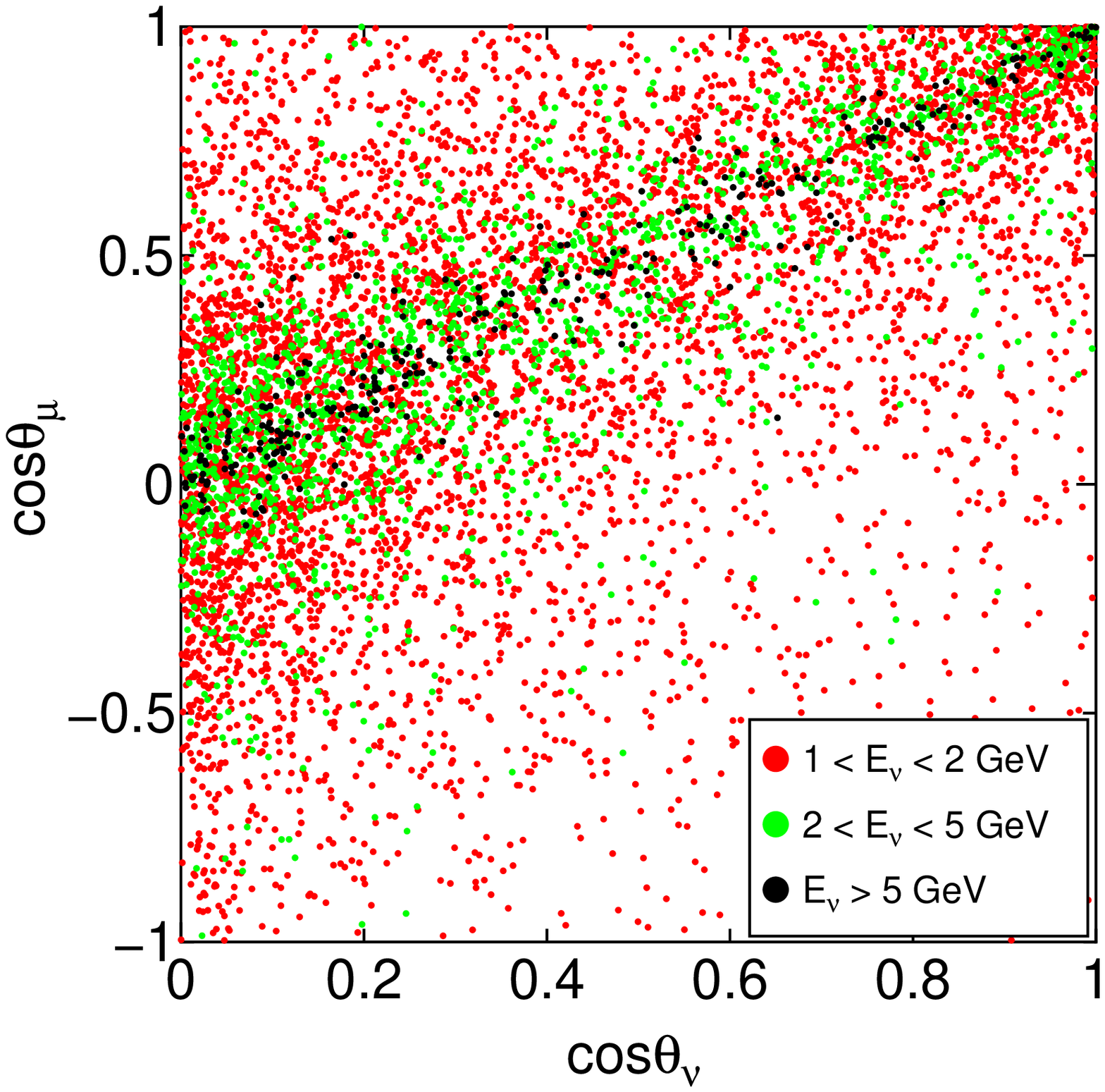}
  }
\caption{\label{fig:11} Correlation Diagram between $\cos{\theta}_{\nu}$ and $\cos{\theta}_{\mu}$ 
for different neutrino energy regions.}
\end{center}
\end{figure} 

\begin{figure*}
\begin{center}
\resizebox{0.9\textwidth}{!}{%
\includegraphics{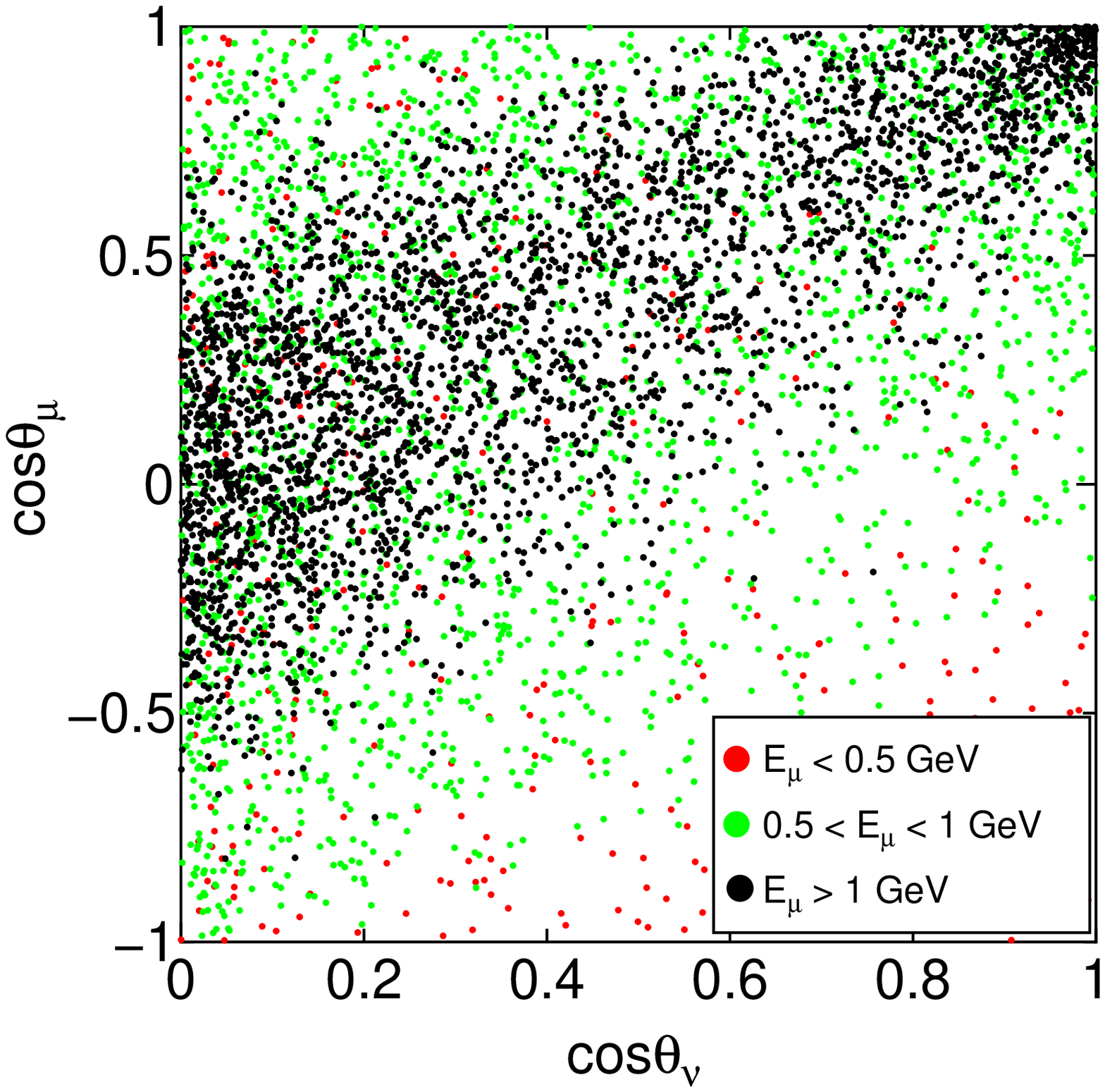}
\hspace{1cm}
\includegraphics{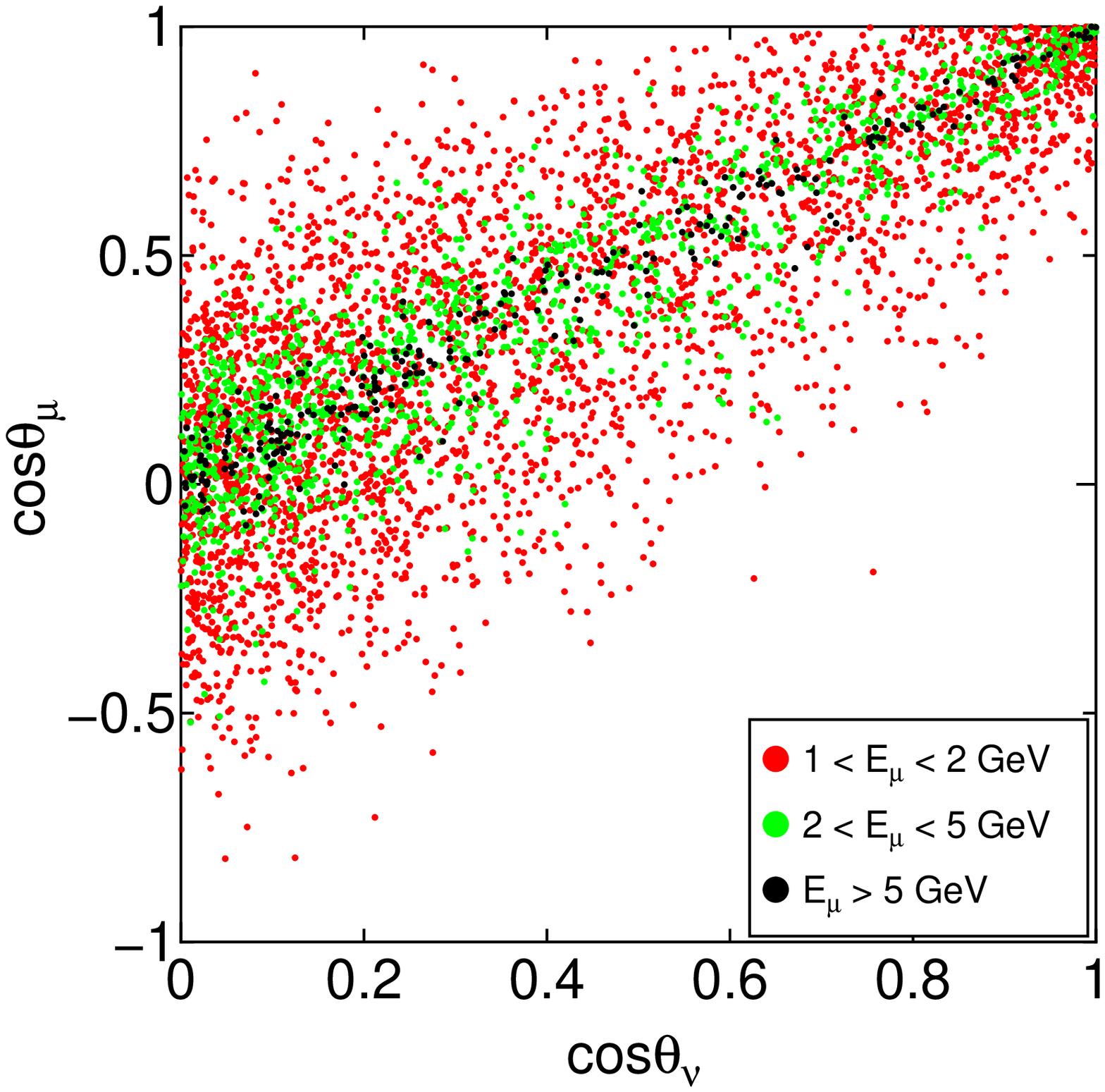}
}
\caption{\label{fig:12} Correlation diagrams between ${\cos\theta_{\nu}}$ and $\cos{\theta}_{\mu}$ 
for different muon energy ranges.}
\end{center}
\end{figure*}

In Fig. \ref{fig:11} we classsify the correlation between ${\cos\theta_{\nu}}$ and ${\cos{\theta}_{\mu}}$ according to the different energy range of the incident muon neutrinos. It should be noticed that the SK assumption on ${\cos\theta_{\nu}}$ = ${\cos{\theta}_{\mu}}$ is 
roughly hold only for  $E_{\nu} \geq 5$  GeV, but the widths in ${\cos{\theta}_{\mu}}$ for the definite ${\cos\theta_{\nu}}$ near ${\cos\theta_{\nu}}$ =0 (for horizontally incident neutrino ) are much larger than those near ${\cos\theta_{\nu}}$ =1 ( for the vertically incident neutrino). Of course, this is due to the effect of the azimutal angle in QEL
which  could not be derived by the SK simulation (\textit{DETECTOR SIMULATION}). Such tendencies become more remarkable in  $E_{\nu} \leq 5$ GeV and in these energies the SK assumption on the direction does not hold any more.
\newline
In Fig. \ref{fig:12}, we classify the correlation between ${\cos\theta_{\nu}}$ and ${\cos\theta_{\mu}}$ according to the different energy range of $E_{\mu}$. The similar argument on Fig. \ref{fig:11} can be done on the case of Fig.\ref{fig:12}

Thus, it could be surely concluded from Fig. \ref{fig:11} and Fig. \ref{fig:12}  that the SK assumption on the direction never holds as a good estimator for the determination of the directions of the incident neutrinos.
 
In order to obtain the zenith angle distribution of the emitted leptons for that of the incident neutrinos, we divide the cosine of the zenith angle of the incident neutrino into twenty regular intervals from $\cos{\theta_{\nu}}=0$ to $\cos{\theta_{\nu}}=1$. For the given interval of $\cos{\theta_{\nu}}$, we carry out the exact Monte Carlo simulation, the detail of which is give in the Appendix D and obtain the cosine of the zenith angle of the emitted leptons, taking account of the geometry for surronding the SK detector.
 
Thus, for each interval of $\cos{\theta_{\nu}}$, we obtain the corresponding zenith angle distribution of the emitted leptons. Then, we sum up these corresponding ones over all zenth angles of the incident neutrinos and we finally obtain the relation between the zenith angle distribution for the incident neutrinos and that for the emitted leptons. 

In a similar manner, we could obtain between $\cos{\theta_{\bar{\nu}}}$ and $\cos{\theta_{\bar{\mu}}}$ for anti-neutrinos. The situation for anti-neutrinos is essentially same as that for neutrinos.

Here, we examine the zenith angle distribution of the muons from both upward neutrinos and downward ones in the case that neutrino oscillation does not exist. 

\begin{figure}
\begin{center}
\resizebox{0.45\textwidth}{!}{%
  \includegraphics{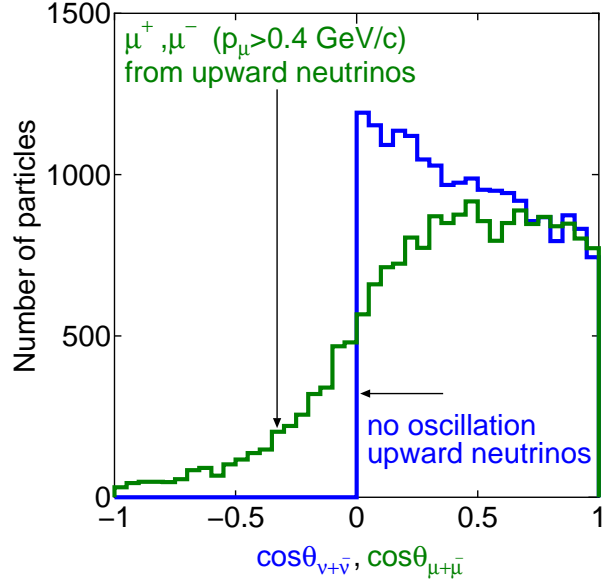}
  }
\caption{\label{fig:13} The relation between the zenith angle distribution of the incident neutrino and
corresponding ones of the emitted lepton}
\end{center}
\end{figure}

By performing the procedures described in Appendix C, a pair of sampling ( $\cos{\theta_{\nu+\bar{\nu}}}$, $E_{\nu+\bar{\nu}}$ ) gives a pair of ( $\cos{\theta_{\mu+\bar{\mu}}}$, $E_{\mu+\bar{\mu}}$ ). In Fig. \ref{fig:13}, we give the zenith angle distribution of the upward neutrinos ( the sum of $\nu_{\mu}$ and $\bar{\nu}_{\mu}$ ) which is constructed from the energy spectra for different $\cos{\theta_{\nu+\bar{\nu}}}$. (see, Honda et. al.  \cite{honda} and Appendix B) 

Upward neutrinos may produce even downward leptons due to both the backscattering effect and the effect of azimuthal angle on larger forward scattering for the interaction concerned (see, Figure 3 and Figures 4 to 6 in the text). As the result of it, the zenith angle distribution of the emitted muons for the upward neutrino may leak in the downward direction. From Figure 13, it is very clear that the shape of the zenith angle distribution for the incident neutrinos is quite different from that of the emitted muons produced by these neutrinos. If \textit{the SK assumption on the direction} statistically holds, the zenith angle distribution for the emitted muons should coincide totally with that of the incident neutrinos. In other words, one may say that the zenith angle distribution for the emitted muons should be understood as that of the incident neutrino under \textit{the SK assumption on the direction}. However, the muon spectrum is distinctively different from  the real (computational) incident neutrino spectrum as shown in the figure. 
Thus we conclude that SK assumption on the direction leads to the wrong conclusion on the neutrino oscillation.
The further examination on the experimental data obtained by SK will be carried out in the subsequent papers.

It is, further, noticed that upward neutrino energy spectrum in the figure biggest near $\cos{\theta_{\nu+\bar{\nu}}}=0$ and the smallest near $\cos{\theta_{\nu+\bar{\nu}}}=1$, which reflects from the enhancement of the primary incident neutrino energy spectrum from the inclined direction and is independent on the neutrino oscillation, while in SK opinion, such tendency may be favor of the existence of neutrino oscillation.

\section{Discussions and Conclusion}

In order to extract the definite conclusion on the neutrino oscillation from the experiment by cosmic ray neutrinos whose intensity as well as interaction with the substance are both very weak, 
first of all, one should analyze the most clear cut and ambiguity free events.  Among neutrino events analyzed by SK, the most clear cut events are {\it Single Ring Events}, such as electron-like events and muon-like events in {\it Fully Contained Events} which are generated by QEL.
These events are provided with simplicity
due to single ring and all possible measureable physical quantities are confined in the detector. 
  
Furthermore, QEL is the most dominant source for neutrino events which are generated in the SK detector.
This is the reason why we examine the QEL events exculsively in present and subsequent papers.
   
If the neutrino oscillation  really exists, the most clear cut evidence surely appears in the analysis of single ring events due to QEL in \textit{Fully Contained Events} and one does not need the analysis of any other type of events, such as single ring events in \textit{Partially Contained Events}, multi-ring events in either \textit{Fully Contained Events} or \textit{Partially Contained Evetns}, all of which include inevitably ambiguities for the interpretation and show merely sub-evidences compared with that from the single ring events due to QEL in \textit{Fully Contained Events}.

SK analyze the zenith angle distribution of the incident neutrinos under the asssumption that the direction 
of the incident neutrino is the same as that of the emitted lepton.  We conclude that this assumption is supplemented 
by their Monte Carlo Simulation named as {\it Detector  Simulation}
\footnote{The SK {\it Detector Simulation} for obtaing the zenith angle distribtuion of the incident neutrino as for the neutrino oscillatiom has never been disclosed in their papers, even in the Ph.D thesis. Consequently, this is only our onjecture as for utilization of the {\it SK Detector Simulation }. A clear thing is only that SK impose the proposittion that the direction of the incident neutrino is the same as that of the emitted lepton upon the neutrino oscillation analysis.}.

In the present paper, we adopt {\it Time Sequential Simulation} which starts from the incident neutrino energy spectrum 
on the opposite side of the Earth to the SK detector and simulate all posible physical processes which are connected with the zenith angle distribution of the incident neutrinos according to their probability functions concerned 
for the examination on the validity of the SK assumption on the direction.

 Concretely speaking, we take the following treatment, (i) the stochastic treatments of the scattering angle of the emitted lepton in QEL, including the scattering on the backward as well as the azimuthal angle, which could not be treated in {\it Detector Simulation}, (ii) the stochastic treatment on the zenith angle distribution of the emitted lepton, considering the incident neutrino energy spectrum, (iii) the stochastic treatment on the detection of the QEL events inside the SK detector.

Furthermore, the discrimination between \textit{Fully Contained Events} and \textit{Partially Contained Events} is only possible in the \textit{Time Sequential Simulation}, because the events concerned may be classified into different categories by chance, \textit{Fully Contained Events} and \textit{Partially Contained Events} due to different occurring points and different directions.

The conclusions thus obtained  are as follows:

\begin{enumerate}
\renewcommand{\labelenumi}{(\arabic{enumi})}
\item The zenith angle distributions of the emitted lepton in QEL for the incident neutrino with both the definite zenith angle and the definite energy are widely spread, particularly, into even the backward region due to partly pure backscattering and partly the combination of the azimuthal angle with the slant direction of the incident neutrinos.
However, for every incident neutrino with a definte zenith angle, SK give the same definite zenith angle to the emitted lepton. Already in this stage, \textit{the SK assumption on the direction} does not hold.\\

\item Taking account of the incident neutrino energy spectrum and simulating all physical processes concerned, we obtain the zenith angle distribution of the emitted leptons for the incident neutrino with a definite zenith angle. It is proved that \textit{the SK assumption on the direction} does not hold again.\\

\item The correlation diagrams between ${\cos\theta_{\nu}}$ and  ${\cos{\theta}_{\mu}}$ show that SK assumption does noty hold well even for higher energies of the incident neutrinos, and it is shown that the correlation between them become weaker in more inclined incident neutrinos due to the effect of the azimuthal angle in QEL.\\

\item Taking into account the detection efficiency for the events concerned in the simulation for upward neutrinos and anti-neutrinos, we obtain the zenith angle distribution of the leptons ( muons plus anti-muons ). According to the {\it SK Assumption on the direction}, the zenith angle distribution is the same as that of the incident neutrinos. However, the original zenith angle distribution of incident neutrino is found to be quite different from that derived from that of leptons. This is the final conclusion that SK have not measured the direction of the incident neutrinos reliably, which is quite independent on either the existence or non-existence of the neutrino oscillation.\\

\item The SK assume that the {\it Partially Contained Events } exclusively belong to the muon-like event. However, 
such the assumption lacks in theoretical background. 
Electron events can also contribute to the \textit{Partially Contained Events} under some geometrical condition,
 for example, partly coming from the transformation by Eq.(A.5). 
The quantitative examination  on the {\it Partially Contained Events } among the electron-like event will be published elsewhere.   

\end{enumerate}

In subsequent papers, we will give the relation between the zenith angle distributions of the incident neutrinos
and the corresponding muons in the cases with and without neutrino oscillation, including downward neutrino and will
examine whether it is possible to or not to detect the neutrino oscillation by using atmospheric neutrino.
               
\newpage
In the following Appendices we give the concrete Monte Carlo Simulations, namely, the details of our \textit{Time Sequential Simulation}.  
\appendix

\section{\appendixname:{} Monte Carlo Procedure for the Decision of Emitted Energies of the Leptons and Their Direction Cosines }
\setcounter {equation} {0}
\def\theequation{\Alph{section}\textperiodcentered\arabic{equation}}

Here, we give the Monte Carlo Simulation procedure for obtaining the energy and its  direction cosines,
$(l_{r},m_{r},n_{r})$, of the emitted lepton in QEL for a given energy and its direction cosines, $(l,m,n)$, of the incident neutrino. 

The relation among $Q^2$, $E_{\nu+\bar{\nu}}$, the energy of the incident neutrino, $E_{\ell}$, the energy of the emitted lepton (muon or electron or their anti-particles) and $\theta_{\rm s}$, the scattering angle of the emitted lepton, is given as
      \begin{equation}
         Q^2 = 2E_{\nu(\bar{\nu})}E_{\ell(\bar{\ell})}(1-{\rm cos}\theta_{\rm s}).
\label{eqn:a1}  
      \end{equation}
\noindent Also, the energy of the emitted lepton is given by
      \begin{equation}
         E_{\ell(\bar{\ell})} = E_{\nu(\bar{\nu})} - \frac{Q^2}{2M}.
\label{eqn:a2}  
      \end{equation}

\begin{figure}
\begin{center}
\resizebox{0.45\textwidth}{!}{%
  \includegraphics{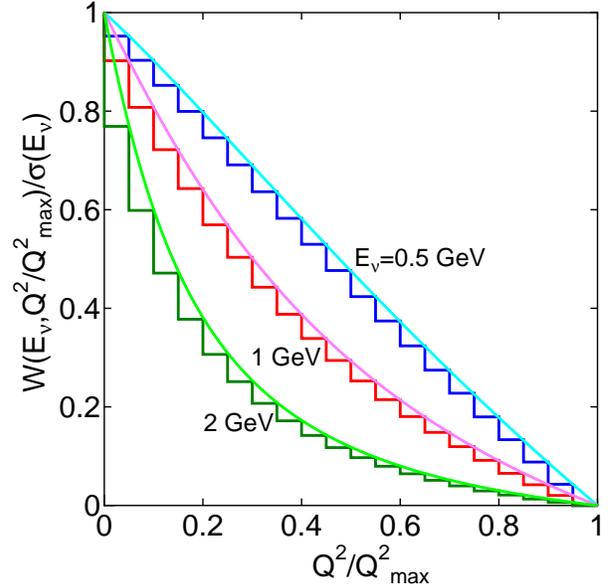}
  }
\end{center}
\caption{\label{fig:a1} The reappearance of the probability function for QEL cross section.
Histograms are  sampling results, while the curves  concerned are theoretical
ones for given incident energies.
}
\end{figure}

\noindent {\bf Procedure 1}\\
\noindent
We decide  $Q^2$ from the probability function for the differential cross section with a given $E_{\nu(\bar{\nu})}$ (Eq. (\ref{eqn:2}) in the text) by using the uniform random number, ${\xi}$,  between (0,1) in the following\\
  \begin{equation}
    \xi = \int_{Q_{\rm min}^2}^{Q^2}P_{\ell(\bar{\ell})}(E_{\nu(\bar{\nu})},Q^2)
                             {\rm d}Q^2,
\label{eqn:a3}  
  \end{equation}
\noindent where
  \begin{eqnarray}
\lefteqn{     P_{\ell(\bar{\ell})}(E_{\nu(\bar{\nu})},Q^2) =} \nonumber \\
&&  \frac{ {\rm d}\sigma_{\ell(\bar{\ell})}(E_{\nu(\bar{\nu})},Q^2) }{{\rm d}Q^2} 
                     \Bigg /\!\!\!\!
      \int_{Q_{\rm min}^2}^{Q_{\rm max}^2} 
      \frac{ {\rm d}\sigma_{\ell(\bar{\ell})}(E_{\nu(\bar{\nu})},Q^2) }{{\rm d}Q^2} 
             {\rm d}Q^2 . \nonumber \\
&&
\label{eqn:a4}  
   \end{eqnarray}
\\
\noindent From Eq. (A$\cdot$1), we obtain $Q^2$ in histograms together with the corresponding theoretical curve in Fig. \ref{fig:a1}. The agreement between the sampling data and the theoretical curve is excellent, which shows the validity of the utlized  procedure in Eq. (A$\cdot$3) is right. \\

\noindent {\bf Procedure 2}\\
\noindent
We obtain $E_{\ell(\bar{\ell})}$ from Eq. (A$\cdot$2) for  the given $E_{\nu(\bar{\nu})}$ and $Q^2$ thus decided in the Procedure 1.\\

\noindent {\bf Procedure 3}\\
\noindent
We obtain $\cos{\theta_{\rm s}}$, cosine of the the scattering angle of the emitted lepton, for $E_{\ell(\bar{\ell})}$ thus decided in the Procedure 2 from Eq. (A$\cdot$1) .\\

\noindent {\bf Procedure 4}\\
\noindent
We decide $\phi$, the azimuthal angle of the scattering lepton, which is obtained from\\
  \begin{equation}
       \phi = 2\pi\xi.                 
\label{eqn:a5}  
  \end{equation}

\noindent Here, $\xi$ is a uniform random number (0, 1). \\
As explained schematically in the text(see Fig. {\bf \ref{fig:3}} in the text),  we must take account of the effect due to the azimuthal angle $\phi$ in the QEL to obtain the zenith angle distribution of both {\it Fully Contained Events} and {\it Partially Contained Events} correctly.\\  

\noindent {\bf Procedure 5}\\
\noindent
The relation between direction cosines of the incident neutrinos, $(\ell_{\nu(\bar{\nu})}, m_{\nu(\bar{\nu})}, n_{\nu(\bar{\nu})} )$, and those of the corresponding emitted lepton, $(\ell_{\rm r}, m_{\rm r}, n_{\rm r})$, for a certain $\theta_{\rm s}$ and $\phi$ is given as \\

\begin{equation}
\left(
         \begin{array}{c}
             \ell_{\rm r} \\
             m_{\rm r} \\
             n_{\rm r}
         \end{array}
       \right)
           =
       \left(
         \begin{array}{ccc}
            \displaystyle \frac{\ell n}{\sqrt{\ell^2+m^2}} & 
            -\displaystyle 
            \frac{m}{\sqrt{\ell^2+m^2}}        & \ell_{\nu(\bar{\nu})} \\
            \displaystyle \frac{mn}{\sqrt{\ell^2+m^2}} & \displaystyle 
            \frac{\ell}{\sqrt{\ell^2+m^2}}     & m_{\nu(\bar{\nu})}    \\
                        -\sqrt{\ell^2+m^2} & 0 & n_{\nu(\bar{\nu})}
         \end{array}
       \right)
       \left(
          \begin{array}{c}
            {\rm sin}\theta_{\rm s}{\rm cos}\phi \\
            {\rm sin}\theta_{\rm s}{\rm sin}\phi \\
            {\rm cos}\theta_{\rm s},
          \end{array}
       \right),
\label{eqn:a6}
\end{equation}
\\
\noindent where $n_{\nu(\bar{\nu})}={\rm cos}\theta_{\nu(\bar{\nu})}$, and $n_{\rm r}={\rm cos}\theta_{\ell}$. 
Here, $\theta_{\ell}$ is the zenith angle of the emitted lepton. \\

\begin{figure}
\begin{center}
\resizebox{0.45\textwidth}{!}{%
  \includegraphics{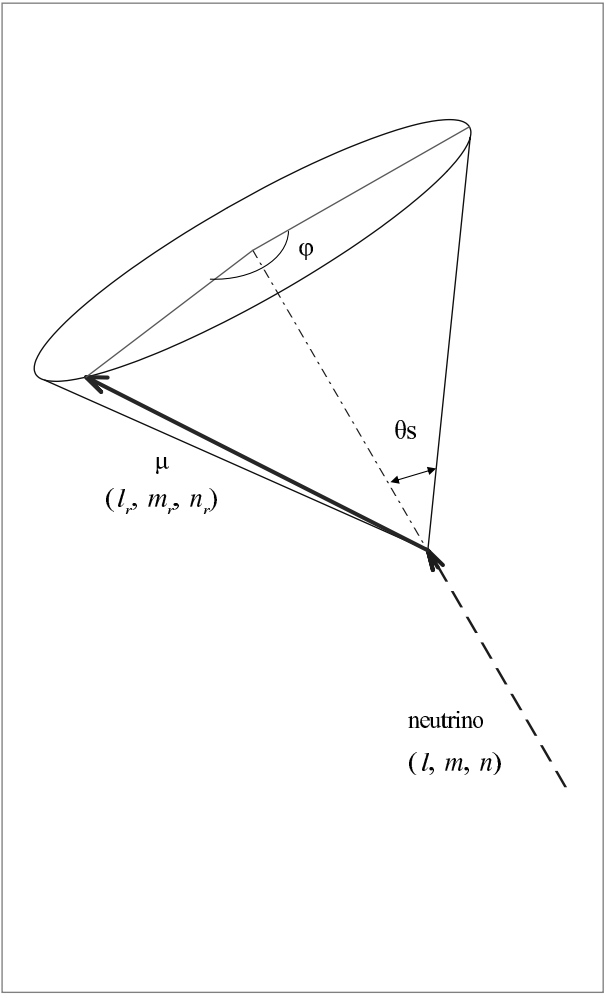}
  }
\end{center}
\caption{\label{fig:14} The relation between the direction cosine of the incident neutrino and that of the emitted charged lepton.}
\end{figure} 

The Monte Carlo procedure for the determination of $\theta_{\ell}$ of the emitted lepton for the parent (anti-)neutrino with given $\theta_{\nu(\bar{\nu})}$ and $E_{\nu(\bar{\nu})}$ involves the following steps:\\

We obtain $(\ell_r, m_r, n_r)$ by using Eq. (\ref{eqn:a6}). The $n_r$ is the cosine of the zenith angle of the emitted lepton which should be contrasted to $n_{\nu}$, that of the incident neutrino.
\\
Repeating the procedures 1 to 5 just mentioned above, we obtain the zenith angle distribution of the emitted leptons for a given zenth angle of the incident neutrino with a definite energy. \\

In the SK analysis,  instead of Eq. (\ref{eqn:a6}), they assume \\ $n_r = n_{\nu(\bar{\nu})} $ 
uniquely for ${E_{\mu(\bar{\mu})}} \geq$ 400 MeV.\\

\newpage
\section{\appendixname:{ Monte Carlo Procedure to Obtain the Zenith Angle of the Emitted Lepton for a Given Zentith Angle of the Incident Neutrino}}
 \setcounter {equation}{0}
 \def\theequation{\Alph{section}\textperiodcentered\arabic{equation}}

  The present simulation procedure for a given zenith angle of the incident neutrino starts from the atmospheric neutrino spectrum at the opposite site of the Earth to the SK detector. We define, $N_{\rm int}(E_{\nu(\bar{\nu})},t,{\rm cos}\theta_{\nu(\bar{\nu})})$, the interaction neutrino spectrum at the depth $t$ from the SK detector in the following way
   \begin{eqnarray}
    \lefteqn{  N_{\rm int}(E_{\nu(\bar{\nu})},t,{\cos}\theta_{\nu(\bar{\nu})}) =N_{\rm sp}(E_{\nu(\bar{\nu})},\cos\theta_{\nu(\bar{\nu})}) \times } \nonumber \\
&&  \Bigg(1-\frac{{\rm d}t}{\lambda_1(E_{\nu(\bar{\nu})},t_1,\rho_1)} \Bigg)  \times \cdots \times \Bigg(1-\frac{{\rm d}t}{\lambda_n(E_{\nu(\bar{\nu})},t_n,\rho_n)} \Bigg).\nonumber \\
&& 
\label{eqn:b1}
   \end{eqnarray}

Here, $N_{\rm sp}(E_{\nu(\bar{\nu})},\cos\theta_{\nu(\bar{\nu})})$ is the atmospheric (anti-)  neutrino spectrum for the zenith angle at the opposite surface of the Earth.

%

Here $\lambda_i(E_{\nu(\bar{\nu})},t_i,\rho_i)$ denotes the mean free path due to the neutrino(anti neutrino) with the energy $E_{\nu(\bar{\nu})}$ from QEL at the distance, $t_i$, from the opposite surface of the Earth inside whose density is $\rho_i$. 


The procedures of the Monte Carlo Simulation for the incident neutrino(anti neutrino) with a given energy, $E_{\nu(\bar{\nu})}$, whose incident direction is expressde by $(l,m,n)$ are as follows.\\

\noindent {\bf Procedure A}\\
\noindent
For the given zenith angle of the incident neutrino, ${\theta_{\nu(\bar{\nu})}}$, we formulate, $N_{\rm pro}( E_{\nu(\bar{\nu})},t,\cos\theta_{\nu(\bar{\nu})}){\rm d}E_{\nu(\bar{\nu})}$, the production function for the neutrino flux to produce leptons at the Kamioka site in the following
   \begin{eqnarray}
\lefteqn{N_{\rm pro}( E_{\nu(\bar{\nu})},t,\cos\theta_{\nu(\bar{\nu})}){\rm d}E_{\nu(\bar{\nu})} } \nonumber \\
&&=
 \sigma_{\ell(\bar{\ell})}(E_{\nu(\bar{\nu})}) N_{\rm int}(E_{\nu(\bar{\nu})},t,{\rm cos}\theta_{\nu(\bar{\nu})}){\rm d}E_{\nu(\bar{\nu})},
\label{eqn:b2}
   \end{eqnarray}
%
\noindent where
  \begin{equation}
     \displaystyle
      \sigma_{\ell(\bar{\ell})}(E_{\nu(\bar{\nu})}) = \int^{Q_{\rm max}^2}_{Q_{\rm min}^2}  \frac{ {\rm d}\sigma_{\ell(\bar{\ell})}(E_{\nu(\bar{\nu})},Q^2)}{{\rm d}Q^2}{\rm d}Q^2.
\label{eqn:b3}
  \end{equation}

\noindent Each differential cross section above is given in Eq. (\ref{eqn:2}) in the text.\\
Utilizing, $\xi$, the uniform random number between (0,1), 
we determine $E_{\nu(\bar{\nu})}$, the energy of the incident neutrino 
in the following sampling procedure\\
    \begin{equation}
       \xi = \int_{E_{\nu(\bar{\nu}),{\rm min}}}^{E_{\nu(\bar{\nu})}}
             P_d(E_{\nu(\bar{\nu})},t,\cos\theta_{\nu(\bar{\nu})}(\bar{\nu})){\rm d}E_{\nu(\bar{\nu})},
    \end{equation}
where
\begin{eqnarray}
\lefteqn{
        P_d(E_{\nu(\bar{\nu})},t,\cos{\theta}_{\nu(\bar{\nu})}){\rm d}E_{\nu(\bar{\nu})} 
        } \nonumber\\
&=& 
\frac{
N_{pro}( E_{\nu(\bar{\nu})},t,\cos{\theta}_{\nu(\bar{\nu})}){\rm d}E_{\nu(\bar{\nu})} 
}
{ \displaystyle \int_{E_{\nu(\bar{\nu}),{\rm min}}}^{E_{\nu(\bar{\nu}),{\rm max}}} 
                       N_{pro}( E_{\nu(\bar{\nu})},t,\cos{\theta}_{\nu(\bar{\nu})}){\rm d}E_{\nu(\bar{\nu})} 
} .
\end{eqnarray}

In our Monte Carlo procedure, \\ 
 the reproduction of, 
$P_d(E_{\nu(\bar{\nu})},t,\cos\theta_{\nu(\bar{\nu})}){\rm d}E_{\nu(\bar{\nu})}$, 
the normalized differential neutrino interaction probability function, is confirmed in the same way as in Eq. (A$\cdot$4). 
\\
\\
%
%
%
%
 
\noindent {\bf Procedure B}\\
\noindent
For the (anti-)neutrino concerned with the energy of $E_{\nu(\bar{\nu})}$, we sample $Q^2$ utlizing $\xi_{3}$, the uniform random number between (0,1). The Procedure B is exactly the same as in the Procedure 1 in the Appendix A. \\

\noindent {\bf Procedure C}\\
\noindent
We decide, ${\theta_{\rm s}}$, the scattering angle of the emitted lepton for given $E_{\nu(\bar{\nu})}$ and $Q^2$. The procedure C is exactly the same as in the combination of Procedures 2 and 3 in the 
 Appendix A. \\

\noindent {\bf Procedure D}\\
\noindent
We randomly sample the azimuthal angle of the charged lepton concerned. The Procedure D is exactly the same as in the Procedure 4 in the Appendix A. \\
 
\noindent {\bf Procedure E}\\
\noindent
We decide the direction cosine of the charged lepton concerned. The Procedure E is exactly the same as in the Procedure 5 in the  Appendix A.\\

 We repeat Procedures A to E until we reach the desired trial number. \\
\\

\newpage
\section{\appendixname:{ }Correlation between the Zenith Angles of the Incident Neutrinos and Those of the Emitted Leptons}
\setcounter{equation}{0}
\def\theequation{\Alph{section}\textperiodcentered\arabic{equation}}

\noindent {\bf Procedure A}\\
By using, $N_{\rm pro}( E_{\nu(\bar{\nu})},t,\cos\theta_{\nu(\bar{\nu})}){\rm d}E_{\nu(\bar{\nu})}$,
which is defined 
in Eq. (\ref{eqn:b2}), 
   
\noindent we define the spectrum for $\cos\theta_{\nu(\bar{\nu})}$  in the following.

\begin{eqnarray}
\lefteqn{      I(\cos\theta_{\nu(\bar{\nu})}){\rm d}(\cos\theta_{\nu(\bar{\nu})}) = } \nonumber \\
&&      {\rm d}(\cos\theta_{\nu(\bar{\nu})})
       \int_{E_{\nu(\bar{\nu}),{\rm min}}}^{E_{\nu(\bar{\nu}),{\rm max}}}
 \!\!\!\!\!\!\!\!\!\!\!\!\!\!\!\!N_{\rm pro}( E_{\nu(\bar{\nu})},t,\cos\theta_{\nu(\bar{\nu})}){\rm d}E_{\nu(\bar{\nu})}.
\label{eqn:c1}
\end{eqnarray}

\noindent By using Eq. (\ref{eqn:c2}) and $\xi$, a sampled uniform random number 
between (0,1), then we could determine $\cos\theta_{\nu(\bar{\nu})}$
from the following equation
    \begin{equation}
      \xi = \int_0^{\cos\theta_{\nu(\bar{\nu})}}P_n(\cos\theta_{\nu(\bar{\nu})})
                          {\rm d}(\cos\theta_{\nu(\bar{\nu})}),
\label{eqn:c2}
    \end{equation}

\noindent where
    \begin{equation}
       P_n(\cos\theta_{\nu(\bar{\nu})}) =  I(\cos\theta_{\nu(\bar{\nu})}) 
           \Bigg/
       \int_0^1 I(\cos\theta_{\nu(\bar{\nu})}){\rm d}(\cos\theta_{\nu(\bar{\nu})}).
\label{eqn:c3}
    \end{equation}
\\
\noindent {\bf Procedure B}\\
\noindent
For the sampled ${\rm d}(\cos\theta_{\nu(\bar{\nu})})$ in the Procedure A, we sample 
$E_{\nu(\bar{\nu})}$ from Eq.(\ref{eqn:c4}) by using ${\xi}$, the uniform randum number between (0,1) 

 \begin{equation}
    \displaystyle
       \xi = \int_{E_{\nu(\bar{\nu}),{\rm min}}}^{E_{\nu(\bar{\nu})}} 
                    P_{pro}(E_{\nu(\bar{\nu})},\cos\theta_{\nu(\bar{\nu})}){\rm d}E_{\nu(\bar{\nu})}, 
\label{eqn:c4}
    \end{equation}
where
      \begin{eqnarray}
\lefteqn{
         P_{pro}(E_{\nu(\bar{\nu})},t,\cos\theta_{\nu(\bar{\nu})}){\rm d}E_{\nu(\bar{\nu})} =
} \nonumber \\
&&  
\frac{         N_{\rm pro}( E_{\nu(\bar{\nu})},t,\cos\theta_{\nu(\bar{\nu})}){\rm d}E_{\nu(\bar{\nu})}
}
{\displaystyle         \int_{E_{\nu(\bar{\nu}),{\rm min}}}^{E_{\nu(\bar{\nu}),{\rm max}}} 
         N_{\rm pro}( E_{\nu(\bar{\nu})},t,\cos\theta_{\nu(\bar{\nu})}){\rm d}E_{\nu(\bar{\nu})}
}.
\label{eqn:c5}
      \end{eqnarray}

\noindent {\bf Procedure C}\\
\noindent 
For the sampled $E_{\nu(\bar{\nu})}$ in the Procedure B, we sample $E_{\mu(\bar{\mu})}$ from Eqs. (\ref{eqn:a2}) and  (\ref{eqn:a3}). For the sampled  $E_{\nu(\bar{\nu})}$ 
and $E_{\mu(\bar{\mu})}$, we determine $\cos{\theta}_s$, the scattering angle of the muon uniquely from Eq. (\ref{eqn:a1}).\\

\noindent {\bf Procedure D}\\
\noindent
We determine, $\phi$, the azimuthal angle of the scattering lepton from Eq. (\ref{eqn:a5}) by using ${\xi}$, an uniform randum number between (0,1). \\

\noindent {\bf Procedure E}\\
\noindent  
We obtain $\cos{\theta}_{\mu(\bar{\mu})}$ from Eq. (\ref{eqn:a6}).  As the result, we obtain a pair of ($\cos\theta_{\nu(\bar{\nu})}$, $\cos{\theta}_{\mu(\bar{\mu})}$) through Procedures A to E. Repeating the Procedures A to E, we finally the correlation between the zenith angle of the incident neutrino and that of the emitted muon. 

\newpage
\section{\appendixname:{ }Detection of the Neutrino Events in the SK 
Detector and Their Interaction Points}
\setcounter{equation}{0}
\def\theequation{\Alph{section}\textperiodcentered\arabic{equation}}

\begin{figure}[b]
\begin{center}
\resizebox{0.45\textwidth}{!}{%
  \includegraphics{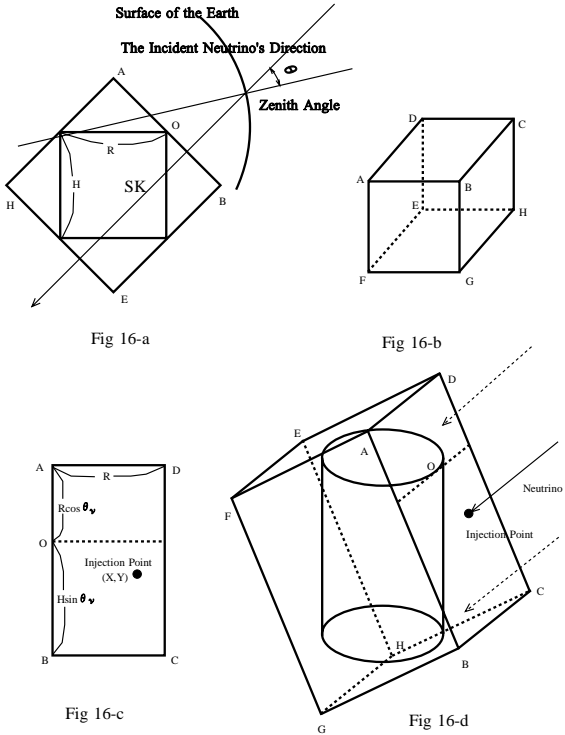}
  }
\end{center}
\caption{\label{fig:ABCD} Sampling procedure for neutrino evens injected into the detector}
\end{figure}

The plane ABCD is always directed vertically to the direction of the incident neutrino with a given zenith angle, which is shown in Fig. \ref{fig:ABCD}. The rectangular ABCDEFGH encloses the SK detector whose radius and height is denoted by \textit{R} and \textit{H}, respectively. The width and the height of the plane ABCD for a given zenith angle,  ${\theta_{\nu(\bar{\nu})}}$, is given as, \textit{R} and
\textit{R}$\cos\theta_{\nu(\bar{\nu})}$ + \textit{H}$\sin{\theta}_{\nu}$,respectively, which are shown in Fig. 16-c.

Now, let us estimate the ratio of the number of the neutrino events inside the 
SK detector to that in the rectangular ABCDEFGH. As the number of the neutrino 
events inside some material is proportional to the number of the nucleons in the material concerned. 
The number of the nucleons inside the SK detector $(\rho=1)$ is given as

     \begin{equation}
         N_{\rm sk}=\frac{\pi}{4}N_{\rm avoga}R^2H,
     \end{equation}
\\
\noindent where $N_{\rm avoga}$ denotes the Avogadro number, and the number of the nucleons in the exterior of the SK detector inside ABCDEFGH is given as
\\
    \begin{eqnarray}
       N_{\rm extr}(\cos\theta_{\nu(\bar{\nu})}&) &=\rho N_{\rm avoga} 
       \biggl[ \biggl( 1-\frac{\pi}{4} \biggr)R^2H + \nonumber \\
       & &      R(H^2 \!\! + \!\! R^2)\sin{\theta}_{\nu(\bar{\nu})}\cos\theta_{\nu(\bar{\nu})} \biggr],
     \end{eqnarray}

\noindent  where $\rho$ is the density of the rock which surrounds the SK detector. 

Then, the total number of the target in the rectangularABCDEFGH is given as
\\
    \begin{equation}
       N_{\rm tot}(\cos\theta_{\nu(\bar{\nu})}) =N_{\rm sk}+N_{\rm extr}(\cos\theta_{\nu(\bar{\nu})}) .
    \end{equation}

\noindent Here, we take 2.65, as ${\rho}$  (standard rock).\\

Then, $R_{\rm theor}$, the ratio of the number of the neutrino events in the SK detector to that in the rectangular ABCDEFGH is given as

   \begin{equation}
          R_{\rm theor}(\cos\theta_{\nu(\bar{\nu})})  = N_{\rm sk} /  N_{\rm tot}(\cos\theta_{\nu(\bar{\nu})})   .
   \end{equation}

\noindent We obtain $R_{\rm theor}$ for different values of $\cos{\theta_{\nu(\bar{\nu})}}$ given in the Table 3.
\\
  Here, we simulate neutrino events occured in the rectangular ABCDEFH, by using the atmospheric neutrino beam which falls down on the plane ABCD. Thus, $N_{\rm smaple}$, the sampling number of the (anti-)neutrino events inside the rectangular ABCDEFG for a given ${\cos\theta_{\nu(\bar{\nu})}}$ 
is given as
\\
   \begin{eqnarray}
\lefteqn{
       N_{\rm sample} (\cos\theta_{\nu(\bar{\nu})}) = N_{\rm tot}(\cos\theta_{\nu(\bar{\nu})})\times 
} \nonumber \\
&&      \int_{E_{\nu(\bar{\nu}),{\rm min}}}^{E_{\nu(\bar{\nu}),{\rm max}}}
       \sigma_{_{\ell(\bar{\ell})}}(E_{\nu(\bar{\nu})})N_{\rm int}(E_{\nu(\bar{\nu})},t,{\rm cos}\theta_{\nu(\bar{\nu})}) 
      {\rm d}E_{\nu(\bar{\nu})} \nonumber \\
      &&
   \end{eqnarray}

\noindent where $\sigma_{\ell(\bar{\ell})}(E_{\nu(\bar{\nu})})$  is the total cross section for (anti-)neutrino due to QEL, and 
$N_{\rm int}(E_{\nu(\bar{\nu})},t,{\rm cos}\theta_{\nu(\bar{\nu})}){\rm d}E_{\nu(\bar{\nu})}$ 
is the differential nutrino energy spectrum for the definite zenith angle, $\theta_{\nu(\bar{\nu})}$, in the plane ABCD. The injection points of the neutrinos in the plane ABCD are distributed over the plane randomly and uniformely and the injection points are determined from a pair of the uniform random  numbers between (0,1). They penetrate into the rectangular ABCDEFGH from the injection point in the plane ABCD and some of them may penetrate into the SK detector or may not, which depend on their injection point.

In the neutrino events which penetrate into the SK detectorr, their geometrical total track length, $T_{\rm track}$, are devided into three parts\\
    \begin{equation}
       T_{\rm track} = T_{\rm b} + T_{\rm sk} + T_{\rm a},
    \end{equation}
\\
\noindent where $T_{\rm b}$ denotes the track length from the plane ABCD to the entrance point of the SK detector, $T_{\rm sk}$ denotes the track length inside the SK detector, and $T_{\rm a}$ denotes the track length from the escaping point of the SK detector to the exit point of the rectangular ABCDEF, 
and thus $T_{\rm track}$ denotes the geometrical length of the neutrino concerned in the rectangular ABCDEFGH. 
\newline

By the definition, the neutrinos concerned with $T_{\rm track}$ interact surely somewhere along the $T_{\rm track}$. Here, we are interested only in the interaction point which ocuurs along $T_{\rm sk}$.  We could determine the interaction point in the $T_{\rm sk}$  in the following.\\

We define the following quantities for the purpose.
   \begin{eqnarray}
         T_{\rm weight} &=& T_{\rm sk} + \rho(T_{\rm b} + T_{\rm a}),\\
         \rho_{\rm av}  &=& T_{\rm weight} / T_{\rm track},\\
         \xi_{\rho} &=& \rho_{\rm av} / \rho,\\                         
         \xi_{\rm sk}   &=& T_{\rm sk} / T_{\rm weight}.
    \end{eqnarray}

\begin{figure}
\begin{center}
\resizebox{0.55\textwidth}{!}{%
  \includegraphics{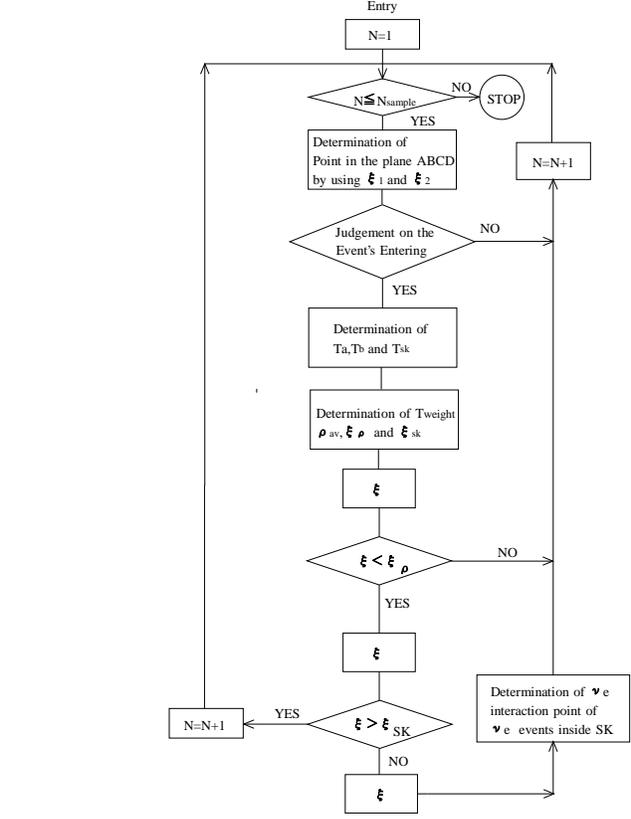}
  }
\end{center}
\vspace{-0mm}
\caption{\label{fig:16} Flow Chart for the determination of the interaction points of the neutrino events 
 inside the detector}
\end{figure}

The flow chart for the choice of the neutrino events in the SK detector and the determination of the interaction points inside the SK detector is given in Fig. \ref{fig:16}. Thus, we obtain neutrino events whose occurrence point is decided in the SK detector in the following.
\
     \begin{eqnarray}
       x_f &=& x_0,\\
       y_f &=& y_0 + \xi T_{\rm sk}\sin{\theta}_{\nu(\bar{\nu})}\\
       z_f &=& z_0 + \xi T_{\rm sk}\cos\theta_{\nu(\bar{\nu})}.
     \end{eqnarray}

If we carry out the Monte Carlo Simulation, following the flow chart in Fig. \ref{fig:16}, then, we obtain $N_{\rm event}$, the number of the neutrino events generasted in the SK detector. The ratio of the selected events to the total trial is given as
   \begin{equation}
          R_{\rm monte}(\cos\theta_{\nu(\bar{\nu})}) = N_{\rm event}(\cos\theta_{\nu(\bar{\nu})}) /  N_{\rm sample}( \cos\theta_{\nu(\bar{\nu})}).
   \end{equation}

\noindent Comparison between $R_{\rm theor}$ and $R_{\rm monte}$ in 
Table 3 shows that our Monte Carlo procedure is valid. 
\begin{table}
\caption{{\label{tab:table3}}Occurrence probabilities of the neutrino events inside the SK detector for different $\cos \theta_{\nu}$'s. Comparison between $R_{\rm theor}$ and $R_{\rm monte}$. The sampling numbers for the Monte Carlo Simulation are, 1000, 10000, 100000, respectively.}
\begin{center}
\begin{tabular}{c c c c c}
\hline
cos $\theta_{\nu}$  & \, $R_{\rm {theor}}$ & \multicolumn{3}{c}{\, \, $R_{\rm {monte}}$} \\ \cline{3-5}
 & & \multicolumn{3}{c}{\, \, Sampling Number} \\ \cline{3-5}
 & & \, \, 1000 & 10000 & 100000 \\
\hline
0.000&\, 0.58002&\, \, 0.576&0.5750&0.57979\\
0.100&\, 0.41717&\, \, 0.425&0.4185&0.41742\\
0.200&\, 0.32792&\, \, 0.353&0.3252&0.32657\\
0.300&\, 0.27324&\, \, 0.282&0.2731&0.27163\\
0.400&\, 0.23778&\, \, 0.223&0.2329&0.23582\\
0.500&\, 0.21491&\, \, 0.206&0.2063&0.21203\\
0.600&\, 0.20117&\, \, 0.197&0.1946&0.19882\\
0.700&\, 0.19587&\, \, 0.193&0.1925&0.19428\\
0.800&\, 0.20117&\, \, 0.198&0.2002&0.20001\\
0.900&\, 0.22843&\, \, 0.230&0.2248&0.22803\\
1.00 &\, 0.58002&\, \, 0.557&0.5744&0.57936\\
\hline
\end{tabular}
\end{center}
\end{table}

%
%

\end{document}